\def\fun#1#2{\lower3.6pt\vbox{\baselineskip0pt\lineskip.9pt
  \ialign{$\mathsurround=0pt#1\hfil##\hfil$\crcr#2\crcr\sim\crcr}}}
\relax
\documentstyle[12pt]{article}
\advance\textheight by \topskip
\begin{document}
\thispagestyle{empty}
\begin{flushright}
SU--ITP--94--36\\
hep-th/9410082\\
\end{flushright}
\hskip 1cm
\vskip 1.5 cm
\begin{center}
 {\Large\bf      LECTURES ON INFLATIONARY COSMOLOGY}

\

 {\bf ANDREI LINDE  \\}

\

{ Physics Department, Stanford University, Stanford, CA 94305\\}
\end{center}
\vglue 0.3cm
\vglue 0.8cm

\begin{center}
{\large CONTENTS}\footnote{Based on the lectures and  talks given in 1994 at
the School  on Particle Physics and Cosmology at Lake Louise, Canada,  at the
Marcel Grossmann Conference, Stanford,  at the Workshop on Birth of the
Universe, Roma,   at the Symposium on Elementary Particle Physics, Capri, and
at the School of Astrophysics,  Erice, Italy}
\end{center}
\begin{quote}

\hskip 3.5 cm1. Introduction

\hskip 3.5 cm 2. The simplest  model of inflation

\hskip 3.5 cm 3. The way inflation begins

\hskip 3.5 cm 4. The way inflation ends. Hybrid inflation

\hskip 3.5 cm 5. Reheating of the Universe after inflation

\hskip 3.5 cm 6. Topological defects as seeds for eternal inflation

\hskip 3.5 cm 7. Stationary Universe

\hskip 3.5 cm 8. Cosmological constant, $\Omega < 1$, and all that

\hskip 3.5 cm 9. Conclusions

\end{quote}

\vfill
\newpage
\section { Introduction} \label{Intro}

The standard big bang theory  asserts that the Universe was born
at some moment $t = 0$ about 15 billion years ago, in a state of
infinitely
large density and temperature. With the rapid expansion  of the
Universe the
average energy
of particles, given by the temperature, decreased rapidly, and the
Universe
became cold.  This theory became especially popular after the
discovery of
the  microwave background radiation. However, by the end of the 70's
it was
understood that this theory is  hardly compatible with the  present
theory of
elementary particles  (primordial monopole problem, Polonyi fields
problem,
gravitino problem,  domain wall problem) and it has many internal
difficulties
(flatness problem,  horizon problem, homogeneity and isotropy
problems, etc.).

Fortunately,  all these problems can be solved simultaneously in the
context of
a relatively simple scenario of the Universe evolution --- the
inflationary
Universe  scenario \cite{b13}--\cite{MyBook}.  The main
idea of
this scenario is that the Universe at the very early stages of its
evolution
expanded quasi-exponentially (the stage of inflation) in a state with
energy
density dominated by the potential energy density $V(\phi)$ of some
scalar
field $\phi$. This rapid expansion made the Universe flat,
homogeneous and
isotropic, and made the density of monopoles,
gravitinos and
domain walls vanishingly small. Later, the potential energy density of the
scalar field
transformed into thermal energy, and  still later, the Universe was
correctly
described by the standard hot Universe theory predicting the
existence of the
microwave background radiation.

The first models of inflation were formulated in the context of the
big bang
theory. Their success  in solving internal problems of this theory
apparently
removed the last doubts concerning the  big bang cosmology.  It
remained almost
unnoticed that during the last ten years   inflationary theory
changed
considerably.  It has broken an umbilical cord connecting it with the
old big
bang theory, and acquired an independent life of its own. For the
practical
purposes of describing   the observable part of our Universe one may
still
speak about the big bang, just as one can still use Newtonian gravity
theory
to describe the Solar system with very high precision. However, if
one tries
to understand the beginning of the Universe, or its end, or its
global
structure, then some of the notions of the big bang theory become
inadequate.
For example, one of the main principles of the big bang theory is the
homogeneity of the Universe. The assertion of homogeneity seemed to
be so
important that it was called  ``the cosmological principle''
\cite{Peebles}. Indeed, without using this principle one could not
prove that
the whole Universe appeared  {\it at a single moment of time}, which
was
associated with the big bang. So far, inflation remains the only
theory which
explains why the observable part of the Universe is almost
homogeneous.
However, almost all versions of inflationary cosmology predict that
on a much
larger scale the Universe should be extremely inhomogeneous, with
energy
density varying from the Planck density to almost zero. Instead of
one single
big bang producing a single-bubble Universe, we are speaking now
about
inflationary bubbles producing new bubbles, producing new bubbles,
{\it ad
infinitum}   \cite{b19,b20}. Thus, recent development of inflationary
theory
considerably modified our cosmological paradigm \cite{MyBook}.  In
order to
understand better this modification, we should remember the main
turning points
in the evolution of the inflationary theory.

The history of inflationary cosmology goes back to 1965, to the papers by Erast
Gliner \cite{b13}. He suggested that the Universe should begin its expansion in
the vacuum-like state, and even calculated the desirable amount of inflation
$\sim e^{70}$. However, at that time almost nobody took his ideas seriously,
except for Andrei Sakharov, who  made an attempt to calculate density
perturbations produced in this scenario   \cite{Sakharov}.

The first semi-realistic version of inflationary cosmology was
suggested
by Alexei Starobinsky \cite{b14}. However, originally it was not quite clear
what
should be the initial state of the Universe in this scenario.
Inflation in
this model could not occur if the Universe was hot from the very
beginning.
To solve this problem, Zeldovich in 1981 suggested that the
inflationary
Starobinsky Universe was created ``from nothing''  \cite{Zeld}. This
idea,
which is very popular now \cite{Creation},  at that time
seemed too
extravagant, and most  cosmologists preferred to study inflation in
a more
traditional context of the hot Universe theory.

One of the most important stages of the development of the
inflationary
cosmology was related to the old inflationary Universe scenario by
Alan Guth
\cite{b15}. His scenario was based on the theory of the  first order
cosmological phase transitions, which was developed in the  mid--70's by David
Kirzhnits and myself  \cite{Kirzhnits}. This scenario was based on three
fundamental
propositions:

1. The Universe initially expands in a state with a very high
temperature,
which leads to the symmetry restoration in the early Universe, $\phi
(T) =
0$, where $\phi$ is some scalar field driving inflation (the inflaton
field).

2. The effective potential $V(\phi, T)$ of the scalar field $\phi$
has a
deep local minimum at $\phi = 0$ even at a very low temperature T. As
a result,
the Universe remains in a supercooled vacuum state $\phi = 0$ (false
vacuum)
for a long time. The energy-momentum tensor of such a state rapidly
becomes
equal to $T_{\mu \nu} = g_{\mu \nu} V(0)$, and the Universe expands
exponentially (inflates) until the false vacuum decays.

3. The decay of the false vacuum proceeds by forming  bubbles
containing the
field $\phi_0$ corresponding to the minimum of the effective
potential
$V(\phi)$. Reheating of the Universe occurs due to the bubble-wall
collisions.

The main idea of the old inflationary Universe scenario was very
simple and
attractive, and the role of the old inflationary scenario in the
development of
modern cosmology was very important.
Unfortunately, as it was
pointed out
by Guth in \cite{b15}, this scenario had a major problem. If the rate
of the
bubble formation is bigger than the speed of the Universe expansion,
then the
phase transition occurs very rapidly and inflation does not take
place. On the
other hand, if the vacuum decay rate is small, then the Universe
after the
phase transition becomes unacceptably inhomogeneous.

All attempts to suggest a successful inflationary Universe scenario
failed
until
cosmologists managed to surmount a certain psychological barrier and
renounce
 the aforementioned assumptions, while retaining the main idea
of ref.\ \cite{b15} that the Universe has undergone inflation during
the early
stages of its evolution. The invention of the new inflationary
Universe
scenario \cite{b16} marked the departure from the assumptions (2),
(3). Later
it was shown that  the assumption (1)  also does not hold in all
realistic
models known so far, for two main reasons. First of all, the time
which is
necessary for the field $\phi$ to roll down to the minimum of
$V(\phi, T)$
is typically  too large, so that  either inflation occurs before the
field rolls
to the minimum of $V(\phi, T)$, or it does not occur at all. On the
other hand,
even if the field $\phi$ occasionally was near the minimum of
$V(\phi, T)$ from
the very beginning, inflation typically starts very late, when
thermal energy
drops from $M^4_p = 1$ down to $V(0, T)$. In all realistic models of
inflation
$V(0, T) < 10^{-10}  $, hence inflation may start in a state with
$\phi =
0$ not earlier than at $t \sim 10^4  $. During such a time a
typical
closed Universe would collapse before the conditions necessary for
inflation
could be realized \cite{MyBook}.

The assumption (1)  was finally given up with the invention of the
chaotic
inflation scenario \cite{b17}. The main idea of this scenario was to
abandon
the  assumption that the Universe from the very beginning was hot,
and that the
initial state of the scalar field should  correspond to a minimum of
its
effective potential. Instead of that,  one should study various
initial
distributions of the scalar field $\phi$, including those which
describe the
scalar field outside of its equilibrium state, and investigate in
which case
the inflationary regime may occur.

In other words, the main idea of chaotic inflation was to remove all
unnecessary restrictions on inflationary models inherited from the
old theory
of the hot big bang. In fact, the first step towards this liberation
was
already made when it was suggested to consider quantum creation of
inflationary
Universe from nothing \cite{Zeld,Creation}. Chaotic inflation
naturally
incorporates this idea \cite{Creation}, but it is more general: inflation
in this
scenario can also appear  under a more traditional assumption of
initial
cosmological singularity.

Thus, the main  idea of chaotic inflation is very simple and
general. One
should study all possible initial conditions without insisting that
the
Universe was in a state of thermal equilibrium, and that the field
$\phi$ was
in the minimum of its effective potential from the very beginning.
However,
this idea   strongly deviated from the standard lore and was
psychologically
difficult to accept. Therefore, for many years almost all
inflationary model
builders continued the investigation of the new inflationary scenario
and
calculated high-temperature effective potentials in all  theories
available. It
was argued that every inflationary model should satisfy the so-called
`thermal
constraints' \cite{TurStein}, that chaotic inflation requires
unnatural initial
conditions, etc.

At present the situation is quite opposite. If anybody ever discusses
the
possibility that inflation is initiated by high-temperature effects,
then
 typically the purpose of such a discussion is to show over and over
again that
this idea  does not work (see
e.g.  \cite{Thermal}), even though some exceptions from this rule
might still
be possible. On the other hand, there are many theories where one can
have
chaotic inflation  (for a review see \cite{MyBook}).

A particularly simple realization of this scenario can be achieved in
the
theory of a massive scalar field with the effective potential
${m^2\over 2}
\phi^2$, or in the theory ${\lambda\over 4}\phi^4$, or in any other
theory with
an effective potential which grows as $\phi^n$ at large $\phi$
(whether or
not there is a spontaneous symmetry breaking at small $\phi$).
Therefore a lot
of work illustrating the basic principles of chaotic inflation was
carried out
in the context of these simple models. However, it would be
absolutely
incorrect to identify the chaotic inflation scenario with these
models, just
as it would be  incorrect to identify new inflation with  the
Coleman-Weinberg
theory.  The dividing line between the new inflation and the chaotic
inflation
was not in the choice of a specific class of potentials, but in the
abandoning
of the idea that the high-temperature phase transitions should be a
necessary
pre-requisite of inflation.

Indeed, already in the first paper where the chaotic inflation was
proposed
\cite{b17}, it was emphasized that this scenario can be implemented
not only in
the theories $\sim \phi^n$, but in {\it any} model where the
effective
potential  is sufficiently flat at some $\phi$. In the second paper
on chaotic
inflation \cite{Chaot2} this scenario was implemented in a  model
with an
effective potential of the same type as those used in the new
inflationary
scenario. It was explained   that the standard scenario based on
high-temperature phase transitions cannot be realized in this theory,
whereas
the chaotic inflation can occur there, either due to the  rolling of
the field
$\phi$ from $\phi > 1$, or due to the rolling down from the local
maximum of
the effective potential at $\phi = 0$. In \cite{ExpChaot} it was
pointed out
that chaotic inflation can be implemented in many theories including
the
theories with exponential potentials $\sim e^{\alpha\phi}$ with
$\alpha <
\sqrt{16\pi}$.
This is precisely the same class of theories which  was used  to obtain the
power  law inflation
\cite{Power}.
The class of models where chaotic inflation can be realized includes
also the
models based on the $SU(5)$ grand unified theory \cite{MyBook}, the
$R^2$
inflation  (modified Starobinsky model) \cite{StChaot},
supergravity-inspired
models with polynomial and non-polynomial potentials \cite{Supergr},
`natural
inflation' \cite{Natural}, `extended inflation'  \cite{b90},
\cite{EtExInf},
`hybrid inflation' \cite{Hybrid}, etc.

Thus, from the point of view of initial conditions, all inflationary models
considered at present belong to the general class of chaotic inflation.  We
will describe the simplest version of chaotic inflation theory in Section 2,
and in Section 3 we will discuss the problem of initial conditions for
inflation.

However, the way inflation begins  is not the only possible
way to classify inflationary models.
Another classification describes various regimes which are
possible
during inflation: quasiexponential inflation, power law inflation,
etc.  This
classification is absolutely independent of the issue of initial
conditions.
Therefore it does not make any sense to compare, say, power law
inflation and
chaotic inflation, and to oppose  them to each other. For example, in
\cite{ExpChaot} it was pointed out that the chaotic inflation
scenario, as
distinct from the new inflationary Universe scenario, can be realized
in the
theories with the effective potential $e^{\alpha\phi}$ for $\alpha
\ll \sqrt{16\pi}$. Meanwhile, in \cite{Power} it was shown that this
inflation is
power law. Thus, the inflationary Universe scenario in the theory
$e^{\alpha\phi}$ describes  chaotic power law inflation.

Finally, the third
classification is related to the way inflation ends.  There are three
different possibilities: slow rollover, the first-order
phase transition and the waterfall. The models of the first class describe
slow
rolling of the
inflaton field $\phi$, which gradually becomes faster and faster. A
particular   model of this type is  chaotic inflation in the theories
$\phi^n$.
The models of the second class should contain at least two scalar
fields,
$\phi$ and $\sigma$. They describe a strongly first-order phase
transition
with bubble production which is triggered by the slow rolling of the
field
$\phi$. One of the popular models of this type is the extended
inflation
scenario \cite{b90}, which is a combination of  the  Brans-Dicke
theory and the old
inflationary scenario.

In the beginning it was assumed that the bubbles formed during the
first-order
phase transition could be a useful ingredient of the theory of the
large scale structure
formation. However, later it was realized that one
should make considerable modifications of the original models in
order to
avoid disastrous consequences of the bubble production. One should consider
Brans-Dicke theories with a specifically adjusted effective potential and with
a non-minimal kinetic term for the Brans-Dicke field, which does not seem
pretty natural. According to the most
recent modification \cite{CrittStein},   the bubble formation happens
only after
the end of inflation.  In this case, the end of inflation occurs as in
the standard
slow-rollover scenario.

The waterfall regime appears in the models of two
interacting scalar fields where  inflation may end  by a rapid
rolling of the
field $\sigma$   triggered by the slow rolling of the
field $\phi$  \cite{Hybrid}.
This regime differs both from the slow-rollover and the first-order
inflation.
By changing parameters of this model  one can
continuously interpolate between these two regimes. Therefore we called this
model ``hybrid inflation''.
Hybrid models of such type may share the best features of the
slow-rollover and the first-order models. We will describe the hybrid inflation
model in Section 4.

One of the most important part of inflationary cosmology is the theory of
reheating of the Universe after inflation. In fact, the word ``reheating'' is
somewhat misleading since in most of the inflationary models which we study at
present the assumption that the Universe was hot before inflation is
unnecessary, at typically it is even harmful. Nevertheless, we are not going to
change the old terminology at this point. It is much more important that the
whole theory of reheating recently has been considerably changed. Previously we
thought that reheating could be described by the standard methods of the theory
of particle decay being applied to the inflaton scalar field driving inflation
\cite{1}. Recently it was understood that in many theories the process of
reheating is much more complicated, and it typically begins with a stage of an
extremely rapid decay of the coherently oscillating scalar field due to a broad
parametric resonance \cite{KLSREH}. Later this resonance becomes narrow, and
finally the old theory of particle decay becomes applicable. However, this old
theory should be applied not to the decay of the inflaton field, but to the
decay of the particles produced at the first stages of explosive reheating,
which we called ``pre-heating''. The corresponding theory will be described in
Section 5.

One of the main purposes of inflationary theory was to solve the primordial
monopole problem. The solution was that inflation makes the distance between
different monopoles exponentially large. However,  recently we have learned
that during inflation monopoles themselves could inflate. Moreover, inflation
of monopoles continues without  end even when inflation
ends in the surrounding space. Therefore primordial monopoles (as well as other
topological defects produced during inflation) can  serve as seeds for the
process of eternal self-reproduction of inflationary universe \cite{LL}. This
issue will be discussed in Section 6.

Eternal inflation supported by monopoles is a very interesting and unexpected
effect.
However, eternal inflation is a very general property of inflationary models,
which does not require any topological defects for its realization. In
particular, this effect exists in the simplest versions of the chaotic
inflation scenario  \cite{b19,b20}: If the Universe contains at
least one inflationary domain of a size of horizon ($h$-region) with
a sufficiently large and homogeneous scalar field $\phi$, then this
domain will
permanently produce new $h$-regions of a similar type. During this
process the
total physical volume of the inflationary Universe (which is
proportional to
the total number of $h$-regions) will grow indefinitely. The process
of
self-reproduction of inflationary domains occurs  not only in the
theories with
the effective potentials growing at large $\phi$ \cite{b19,b20}, but
in some
theories with the effective potentials used in old, new and extended
inflation
scenarios as well \cite{b51,b52,b62,EtExInf}.  However, in the models
with the
potentials growing at large $\phi$ the existence of this effect was
most
unexpected, and it leads to especially interesting cosmological
consequences
\cite{MyBook}.

Self-reproduction of
inflationary domains is responsible for the fundamental stationarity
which is
present in many   inflationary models: properties of the parts of the
Universe
formed in the process of self-reproduction  do not depend on the time
when this
process occurs. We call this property of the inflationary Universe
{\it
local  stationarity}. In addition to it, there may exist  a  stationary
distribution of probability $P_p$ to find a
given field
$\phi$ at a given time in a given physical volume. In this case we will be
speaking of  a {\it global
stationarity} of the inflationary Universe.

Inflationary models are divided into several different groups with respect to
the   self-reproduction and stationarity. In some of them inflation is possible
but there is no self-reproduction. In others self-reproduction and the local
stationarity are present, but there is no global stationarity. Finally, there
exist many models where the distribution $P_p$ is stationary
\cite{LLM}--\cite{BL}. We will consider this issue in Section 7.

The possibility that our Universe is stationary (either locally or globally)
shows that instead of being a fireball created 15 billion  years ago, the
Universe may be a huge expanding and self-reproducing fractal consisting of
infinite number of exponentially large domains containing matter in all its
possible phases. This changes our standard notions about what is natural and
what is not. The new cosmological paradigm appears to be much more interesting
and complicated than what we have expected. In the   Section 8  we will briefly
discuss some unusual nonperturbative effects which  may imply that we should
live in a center of a   bubble, so that the density around us is smaller than
the density at a very large distance from us (i.e. smaller than the critical
density). This possibility is similar (though not quite identical) to the
possibility that we live in an inflationary universe with $\Omega < 1$
\cite{LLM2}.

In all parts of   this paper except for the Section 5 describing reheating we
will use the system of units in which the Planck mass $M_p=1$.
\vskip 0.5cm
When I was finishing writing this paper I was informed about the death of
academician Moisey Alexandrovich Markov.  He was an extremely talented
physicist who made a lot for  the development of particle physics and theory of
gravity, and also for   breaking the artificial isolation of the Soviet physics
from the international scientific community. I dedicate this article to his
memory.

\

\section   {The simplest model of inflation}

We will begin our discussion  with the simplest
  chaotic inflation model \cite{b17} based
on the  theory of a  scalar field $\phi$ minimally coupled to
gravity, with the Lagrangian
\begin{equation}\label{E01}
L =  \frac{G}{16\pi}R + \frac{1}{2} \partial_{\mu} \phi
\partial^{\mu} \phi - V(\phi)  \ .
\end{equation}
Here $G = M^{-2}_p$ is the gravitational constant,  $R$ is the
curvature
scalar, and $V(\phi)$ is the effective potential of the scalar field.
 If the
classical field $\phi$ is sufficiently homogeneous in some domain of
the Universe, then its behavior inside this domain is
governed by the
equations
\begin{equation}\label{E02}
\ddot\phi + 3H\dot\phi = -dV/d\phi \ ,
\end{equation}
\begin{equation}\label{E03}
H^2 + \frac{k}{a^2} = \frac{8\pi}{3}\, \left(\frac{1}{2}
\dot\phi^2 + V(\phi)\right) \ .
\end{equation}
Here $H={\dot a}/a, a(t)$ is the scale factor of the
 Universe, $k=+1, -1,$ or $0$ for a closed, open or
flat Universe, respectively.

The simplest version of the  theory (\ref{E01}) is the theory of
a massive
noninteracting scalar field with the effective potential
$V(\phi)=\frac{1}{2}
m^2\phi^2$, where $m$ is the mass of the scalar field $\phi$, $m
\ll1$\@. If the
field $\phi$ initially is sufficiently large, $\phi \gg 1$, then one
can show that
the functions $\phi(t)$ and $a(t)$ rapidly approach the asymptotic
regime
\begin{equation}\label{E04} \phi(t)=\phi_0 - \frac{m}{2(3\pi)^{1/2}}
t \ , \ \ \ \ \ a(t) = a_0\cdot  \exp\left(2 \pi
\left(\phi^2_0 - \phi^2(t)\right) \right) \ .
\end{equation}
Note that in this regime the second derivative of the scalar field in
eq. (\ref{E02}) and  the term $k/a^2$ in  (\ref{E03})
can be neglected. The last statement  means
that the Universe becomes locally flat.

 According to (\ref{E04}), during the time $\tau \sim \phi/m$ (which is much
greater than $H$ for $\phi \gg 1$) the relative change of the field $\phi$
remains small, the effective potential $V(\phi)$ changes very slowly and the
Universe  expands quasi-exponentially:
$a(t+\Delta t)\sim a(t) \exp
(H\Delta t)$
for $\Delta t \leq \tau$\@. Here
$H = 2 \sqrt \frac{\pi}{3} \, m \phi$.

The regime of the slow rolling of the field $\phi$ and the
quasi-exponential
expansion (inflation) of the Universe ends at $\phi {\
\lower-1.2pt\vbox{\hbox{\rlap{$<$}\lower5pt\vbox{\hbox{$\sim$}}}}\ }\phi_e$. In
the theory
under consideration, $\phi_e \sim
0.2$\@.   At $\phi \leq \phi_e$ the field  $\phi$ oscillates rapidly,
and if this
field interacts with other matter fields (which are not written
explicitly in eq.
(\ref{E01})), its potential energy $V(\phi) \sim      \frac{m^2
\phi_e^2}{2}
 \sim {m^2\over 50} $ is transformed into heat. The
reheating temperature $T_R$ may be of the order
$m^{1/2}$ or somewhat smaller, depending on the
strength of the interaction of the field $\phi$ with other
fields. It is important that $T_R$ does not
depend on the initial value $\phi_0$ of the field $\phi$.
  The only parameter which depends on $\phi_0$ is the scale
factor $a(t)$, which grows $e^{2\pi\phi^2_0}$
times during inflation.

All results obtained above can be easily generalized for the theories
with more
complicated effective potentials. For example, during inflation in
the theories
with $V(\phi)=\frac{\lambda}{n}\, \phi^n$ one has
\begin{equation}\label{E04a}
\phi^{4-n\over 2}(t)=\phi^{4-n\over 2}_0 - \frac{4-n}{2}\,
\sqrt{\frac{n \lambda}{24 \pi}} \ t \ \ \ \ \mbox{for} \ n\not = 4\ ,
\end{equation} \begin{equation}\label{E04a'}
\phi(t)=\phi_0 \ \exp\left(-\sqrt{\lambda\over 6\pi} \, t\right) \ \
\ \ \
 \mbox{for} \ n = 4 \ .
 \end{equation}
Inflation ends at $\phi_e \sim {n\over 4\sqrt{3\pi}}$. The scale factor is
given by
\begin{equation}\label{E05a}
a(t) = a_0 \exp \left({4 \pi\over n}
(\phi^2_0 - \phi^2(t))\right) \ .
\end{equation}

Note, that  in all realistic models of elementary particles
spontaneous
symmetry breaking occurs on a scale which is many orders of magnitude
smaller
than 1 (i.e. smaller than $M_p$).
Therefore all results which we obtained remain valid in all theories
which have
potentials $V(\phi) \sim \frac{\lambda}{n}\, \phi^n$ at $\phi {\
\lower-1.2pt\vbox{\hbox{\rlap{$>$}\lower5pt\vbox{\hbox{$\sim$}}}}\ }1$,
independently of the issue of spontaneous symmetry breaking which may
occur in
such theories at small $\phi$.

As was pointed out in \cite{Power,ExpChaot}, chaotic inflation occurs
as well
in the theories with exponential effective potentials $V(\phi)={V_0}\
{e^{\alpha\phi}}$ for sufficiently small $\alpha$.  In such theories the
Einstein equations and equations for the scalar field have an
exact
solution:
\begin{equation}\label{E04a''}
\phi(t)=\phi_0 - {2\over\alpha}\  \ln {t\over t_0}  \ , \ \ \ \ \ \
a(t) = a_0 \, t^p   \ ,
\end{equation}
where $p =   \frac{16 \, \pi}{\alpha^2}$. Note that here we are dealing with
the power law expansion of the
Universe. It
can be called inflation  if $p \gg 1$, which
implies that
$\alpha \ll \sqrt{16\pi} \sim 7$.

\

\section {   The way inflation begins \label{IniCond}}

It is most important to investigate initial conditions which are required for
inflation. Let us consider first a closed Universe of initial  size $\l \sim 1$
(in Planck
units), which emerges  from the
space-time foam, or from singularity, or from `nothing'  in a state
with  the
Planck
density $\rho \sim 1$. Only starting from this moment, i.e. at $\rho
{\ \lower-1.2pt\vbox{\hbox{\rlap{$<$}\lower5pt\vbox{\hbox{$\sim$}}}}\ }1$,
can we describe this domain as  a {\it classical} Universe.  Thus,
at this
initial moment the sum of the kinetic energy density, gradient energy
density,
and the potential energy density  is of the order unity:\, ${1\over
2}
\dot\phi^2 + {1\over 2} (\partial_i\phi)^2 +V(\phi) \sim 1$.

We wish to emphasize, that there are no {\it a priori} constraints on
the
initial value of the scalar field in this domain, except for the
constraint
${1\over 2} \dot\phi^2 + {1\over 2} (\partial_i\phi)^2 +V(\phi) \sim
1$.  Let
us consider for a moment a theory with $V(\phi) = const$. This theory
is
invariant under the shift  $\phi\to \phi + a$. Therefore, in such a
theory {\it
all} initial values of the homogeneous component of the scalar field
$\phi$ are
equally probable. Note, that this expectation would be incorrect if
the scalar
field should vanish at the boundaries of the original domain. Then
the
constraint ${1\over 2} (\partial_i\phi)^2 {\
\lower-1.2pt\vbox{\hbox{\rlap{$<$}\lower5pt\vbox{\hbox{$\sim$}}}}\ }1$ would
tell us that
the scalar
field cannot be greater than $1$ inside a domain of initial size $1$.
 However,
if the original domain is a closed Universe, then it has no
boundaries.  (We
will discuss a more general case shortly.)

The only constraint on the average amplitude of the field appears if
the
effective potential is not constant, but grows and becomes greater
than the
Planck density at $\phi > \phi_p$, where  $V(\phi_p) = 1$. This
constraint
implies that $\phi {\
\lower-1.2pt\vbox{\hbox{\rlap{$<$}\lower5pt\vbox{\hbox{$\sim$}}}}\ }\phi_p$,
but it does not give any reason to
expect that
$\phi \ll \phi_p$. This suggests that the  typical initial value
$\phi_0$ of
the field $\phi$ in such a  theory is  $\phi_0
\sim  \phi_p$.

Thus, we expect that typical initial conditions correspond to
${1\over 2}
\dot\phi^2 \sim {1\over 2} (\partial_i\phi)^2\sim V(\phi) = O(1)$.
Note that if
by any chance ${1\over 2} \dot\phi^2 + {1\over 2} (\partial_i\phi)^2
{\ \lower-1.2pt\vbox{\hbox{\rlap{$<$}\lower5pt\vbox{\hbox{$\sim$}}}}\ }V(\phi)$
in the domain under consideration, then inflation begins,
and  within
the Planck time the terms  ${1\over 2} \dot\phi^2$ and ${1\over 2}
(\partial_i\phi)^2$ become much smaller than $V(\phi)$, which ensures
continuation of inflation.  It seems therefore that chaotic inflation
occurs
under rather natural initial conditions, if it can begin at $V(\phi)
\sim 1$
\cite{MyBook,b17,ExpChaot}.

The assumption that inflation may begin at a very large $\phi$ has
important
implications. For example, in the theory (\ref{E01}) one has
$\phi_0 \sim  \phi_p \sim m^{-1/2}$.
Let us consider for definiteness a closed Universe of a typical
initial size
$O(1)$. Then, according to (\ref{E04}), the total size of the
Universe after
inflation becomes equal to
\begin{equation} \label{E09}
l \sim  \exp \left(2\pi \phi^2_0\right)
\sim  \exp \left(\frac{2\pi}{m^2}\right) \ .
\end{equation}
For $m\sim 10^{-6}$ (which is necessary to produce
density perturbations $\frac{\delta{\rho}}{\rho} \sim 10^{-5}$, see
below) $l \sim  \exp \left(2\pi 10^{12}\right) > 10^{10^{12}} cm$. Thus,
according to this estimate, the smallest possible domain of the
Universe of
initial size \mbox{$O(M_p^{-1}) \sim 10^{-33} cm$}
 after inflation becomes
much larger
than the size of the observable part of the Universe $\sim 10^{28}
cm$\@. This is
the reason why our part of the Universe looks flat, homogeneous and
isotropic.

In what follows we will return many times to our conclusion that the
most
probable initial value of the scalar field corresponds to  $\phi \sim
\phi_p$.
There were many objections to it. Even though all these objections
were answered many years ago
\cite{MyBook,ExpChaot},  we need to discuss here one of these
objections again.
This is important for a proper understanding of the new picture of
the evolution of the Universe in inflationary cosmology.

Assume that the Universe is not closed but infinite, or at least
extremely
large from the very beginning. (This objection does not apply to the
closed Universe scenario discussed above.) In this case one could
argue that
our expectation that $\phi_0 \sim \phi_p \gg 1$ is not very natural
\cite{KolbTurn}.  Indeed, the conditions  $(\partial_i\phi)^2 {\
\lower-1.2pt\vbox{\hbox{\rlap{$<$}\lower5pt\vbox{\hbox{$\sim$}}}}\ }1$
and
$\phi_0 \sim \phi_p \gg 1$ imply that the field $\phi$ should be of
the same
order of magnitude $\phi \sim \phi_p \gg 1$ on a length scale at
least as large
as  $\phi_p$, which is much larger than the scale of horizon $l\sim
1$ at the
Planck time.  But this is highly improbable, since initially (i.e.,
at the
Planck time ) there should be no correlation between values of the
field
$\phi$ in different regions of the Universe separated from one
another by
distances greater than $1$.  The existence of such correlation would
violate
causality. As it is written in \cite{KolbTurn},  the scalar field
$\phi$ must
be smooth on a scale much greater than the scale of the horizon,
which does not
sound very chaotic.

The answer to this objection is very simple \cite{MyBook,ExpChaot}.
We have
absolutely no reason to expect that the overall energy density $\rho$
simultaneously becomes smaller than the Planck energy density in all
causally
disconnected regions of an infinite Universe, since that would imply
the
existence of an acausal correlation between values of $\rho$ in
different
domains of  Planckian size $l_p
\sim 1$.  Thus, each such domain at the Planck time after its
creation looks like an isolated island of
classical space-time, which emerges from the space-time foam
independently of
other such islands.  During inflation, each of these islands
independently
acquires a size many orders of magnitude larger than the size of the
observable part of the Universe.  A typical initial size of a domain
of
classical space-time with $\rho {\
\lower-1.2pt\vbox{\hbox{\rlap{$<$}\lower5pt\vbox{\hbox{$\sim$}}}}\ }1$ is of
the order of the Planck
length.
Outside each of these domains the condition $\rho {\
\lower-1.2pt\vbox{\hbox{\rlap{$<$}\lower5pt\vbox{\hbox{$\sim$}}}}\ }1$ no
longer
holds, and
there is no correlation between values of the field $\phi$ in
different
disconnected regions of classical space-time of size $1$.  But such
correlation
is not  necessary at all for the realization of the inflationary
Universe
scenario: according to the `no hair' theorem for de Sitter space, a
sufficient
condition for the existence of an inflationary region of the Universe
is that
inflation takes place inside a region whose size is of order
$H^{-1}$. In our
case this condition is satisfied.

We wish to emphasize once again that the confusion  discussed above,
involving
the correlation between values of the field $\phi$ in different
causally
disconnected regions of the Universe, is rooted in the familiar
notion of a
very large Universe that is instantaneously created from a singular
state with
$\rho= \infty$,  and instantaneously passes through a state with the
Planck
density $\rho = 1$.   The lack of justification for such a notion is
the very
essence of the horizon problem.  Now, having disposed of the horizon
problem
with the aid of the inflationary Universe scenario, we can possibly
manage to
familiarize ourselves with a different picture of the Universe. In
this picture
the simultaneous creation of the whole Universe is possible only if
its initial
size is of the order $1$, in which case no long-range correlations
appear.
Initial conditions should be formulated at the Planck time and on the
Planck
scale. Within each Planck-size island of the classical space-time,
the initial
spatial  distribution of the scalar field cannot be very irregular
due to the
constraint  $(\partial_i\phi)^2 {\
\lower-1.2pt\vbox{\hbox{\rlap{$<$}\lower5pt\vbox{\hbox{$\sim$}}}}\ }1$. But
this does not impose any
constraints on the average values of the scalar field $\phi$ in each
of such
domains. One should examine all possible values of the field $\phi$
and check
whether they could lead to inflation.

Now let us analyze  the probability for inflation to occur in the theories with
small $V(\phi)$. There are two for such an event to happen. First of all, the
inflationary universe can appear from nothing in a state with $\phi$
corresponding to the maximum of $V(\phi)$. The probability of such an event can
be estimated by the square of the tunneling wave function of the Universe
\cite{Creation}:
\begin{equation}\label{d2}
P \sim \exp\Bigl(-{3\over 8V(\phi)}\Bigr) \ .
\end{equation}
This probability may be quite high if inflation is possible for $V(\phi) \sim
1$, as in the chaotic inflation scenario. However, in the models where
inflation occurs near the maximum of $V(\phi)$ considerations related to the
amplitude of density perturbations require $V(\phi) < 10^{-10}$, which gives a
rather disappointing result for the probability of inflation, $P \sim
\exp\Bigl(-{10^{10}}\Bigr)$.

Another possibility is that the Universe was large and hot from the very
beginning, and it became inflationary only when its thermal energy density
$\sim T^4$ dropped below $V(\phi)$. Inflation begins at this stage only if the
Universe at this moment was sufficiently homogeneous on a scale  greater than
$\Delta x \sim H^{-1} \sim 1/\sqrt{V(\phi)}$. Suppose for simplicity that the
Universe from the very beginning was dominated by ultrarelativistic matter.
Then its scale factor expanded as $\rho^{-4}$, where $\rho$ is the energy
density at the pre-inflationary stage. Therefore at the Planck time the size of
the part of the Universe which later evolved into inflationary domain was not
$  1/\sqrt{V(\phi)}$, but somewhat smaller: $\Delta x \sim V^{-1/4}(\phi)$.
This whole scenario can work only if at the Planck time the domain of this size
was sufficiently homogeneous, ${\delta \rho \over \rho} \ll 1$. However, at the
Planck time this domain   consisted of $V^{-3/4}(\phi)$ domains of a Planck
size, and energy density in each of them was absolutely uncorrelated with the
energy density in other domains. Therefore {\it a priori} one could expect
changes of density $\delta \rho \sim \rho$ when going from one    causally
disconnected parts of the Universe of a size $M_P^{-1} = 1$ to another. Simple
combinatorial analysis suggests that the probability of formation of  a
reasonably homogeneous part of the Universe of a size  $\Delta x \sim
V^{-1/4}(\phi)$ at the Planck time is suppressed by the exponential factor
\begin{equation}\label{d1}
P \sim \exp\Bigl(-{C\, \over V^{3/4}(\phi)}\Bigr) \ ,
\end{equation}
where $C = O(1)$.
To get a numerical estimate, one can take $V(0) \sim 10^{-10}$. This gives $P
{\ \lower-1.2pt\vbox{\hbox{\rlap{$<$}\lower5pt\vbox{\hbox{$\sim$}}}}\ }
10^{-10^{7}}$.

Note, that the arguments given above \cite{b17,ExpChaot} suggest that
initial conditions for inflation are quite natural only if inflation  begins
as close as possible to the Planck density. These arguments do not give any
support to
the models where inflation is possible only at densities much smaller
than $1$.  Inflationary models of that
type require fine-tuned initial conditions, and they
cannot solve the flatness problem, unless we apply some rather sophisticated
arguments based on the theory of a self-reproducing inflationary universe
\cite{LLM}.

\

\section {  The way inflation ends. Hybrid Inflation}

As we already mentioned in the introduction, there exist three different way
for inflation to end. The first one is by slow rolling, which gradually becomes
faster and faster. This possibility is illustrated by the simplest model
considered in the second section. The second possibility is the first order
phase transition \cite{b90}. However, after many modifications of this
scenario, it gradually became rather complicated. It requires specific
adjustments of the effective potential and of the kinetic terms for the
Brans-Dicke field \cite{CrittStein}. The third possibility is the rapid rolling
of one of the scalar field triggered by the slow rolling of another (waterfall)
\cite {Hybrid}. We will discuss this possibility here.

The simplest model where the waterfall regime can appear is the theory of two
scalar fields with the effective potential
\begin{equation}\label{hybrid}
V(\sigma,\phi) =  {1\over 4\lambda}(M^2-\lambda\sigma^2)^2
+ {m^2\over
2}\phi^2 + {g^2\over 2}\phi^2\sigma^2\ .
\end{equation}
Theories of this type were considered in \cite{KL}--\cite{HBKP}. The
main
difference between the models of refs. \cite{KL}--\cite{HBKP} and our
model is a specific choice of parameters, which allows the existence
of the
waterfall regime mentioned above. There is also another important
difference: we will
assume that the field $\sigma$ in this model is the Higgs field,
which  remains as
a physical degree of freedom after the
Higgs effect in an underlying gauge theory with spontaneous symmetry
breaking. This  field acquires only positive values, which
removes the possibility of domain wall formation in this theory.
Usually
it is rather dangerous to take the inflaton field interacting with
gauge fields, since its effective coupling constant $\lambda$ may
acquire large radiative corrections $\sim e^4$, where $e$ is the
gauge coupling constant. In our case this  problem does not appear
since density perturbations in our model remain small at rather large
$\lambda$, see below.

The effective
mass squared of the field $\sigma$ is equal to  $-M^2 + g^2\phi^2$.
Therefore
for $\phi > \phi_c = M/g$ the only   minimum of the effective
potential
$V(\sigma,\phi)$ is at $\sigma = 0$. The curvature of the effective
potential in the $\sigma$-direction is much greater than in the
$\phi$-direction. Thus we expect that at the first stages of
expansion of the Universe the field $\sigma$ rolled down to $\sigma =
0$, whereas the field $\phi$ could remain large for a much longer
time.  For this reason we will consider the stage of inflation at
large $\phi$, with $\sigma = 0$.

At the moment when the inflaton field
$\phi$ becomes smaller than  $\phi_c = M/g$, the phase transition
with the
symmetry breaking occurs. If $m^2 \phi_c^2 = m^2M^2/g^2 \ll
M^4/\lambda$, the Hubble constant at the time of the phase transition
is given by
\begin{equation}\label{2}
H^2 = {2\pi M^4 \over 3 \lambda  } \ .
\end{equation}
Thus we will assume that $M^2 \gg {\lambda m^2\over g^2}$.   We will
assume also that
$m^2 \ll H^2$, which gives
\begin{equation}\label{2a}
M^2\gg m\sqrt{3\lambda  \over 2\pi} \ .
\end{equation}

One can easily verify, that, under this condition, the Universe at
$\phi > \phi_c$ undergoes a stage of inflation. In fact, inflation in
this model occurs even if $m^2$ is somewhat greater than $H^2$. Note
that
inflation at its last stages is driven not by the energy density of
the inflaton
field $\phi$ but by the vacuum energy density $V(0,0) = {M^4\over
4\lambda}$, as in the new inflationary Universe scenario. This was
the reason why we called this model `hybrid inflation' in
\cite{Hybrid}.

Let us study the behavior of the fields $\phi$ and $\sigma$ after the
time
$\Delta t = H^{-1} = \sqrt{3\lambda\over 2\pi M^2}$ from
the
moment $t_c$ when the field $\phi$ becomes equal to $\phi_c$. The
equation of motion of the field $\phi$ during inflation is
$3H\dot\phi = m^2\phi$.
Therefore during the time interval $\Delta t = H^{-1}$ the field
$\phi$
decreases from $\phi_c$ by $\Delta\phi = {m^2\phi_c\over 3 H^2}=
{\lambda m^2 \over 2\pi g M^3}$. The absolute value of the
negative effective mass squared $-M^2 + g^2\phi^2$ of the field
$\sigma$ at that time becomes equal to
\begin{equation}\label{3}
M^2(\phi) =  {\lambda m^2 \over \pi  M^2} \ .
\end{equation}
The  value of $M^2(\phi)$ is much greater than $H^2$ for
$M^3 \ll    \lambda   \ m  $.
In this case the field $\sigma$ within the time $\Delta t \sim
H^{-1}$
rolls down to its minimum  at $\sigma(\phi) = M(\phi)/\sqrt\lambda$,
rapidly oscillates near it and loses its energy due to the expansion
of the
Universe.  However, the field cannot simply relax near this minimum,
since
the effective potential $V(\phi,\sigma)$ at $\sigma(\phi)$ has a
nonvanishing partial derivative
\begin{equation}\label{5}
 {\partial V\over \partial\phi} = {m^2\phi} + {g^2\phi M^2(\phi)\over
\lambda}  \ .
\end{equation}
One can easily check that the motion in this direction becomes very
fast and the field $\phi$ rolls to the minimum of its effective
potential within the time much smaller than $H^{-1}$  if
$M^3 \ll  {\sqrt \lambda\, g } \ m   $.
Thus, under the specified conditions inflation ends in this theory
almost
instantaneously, as soon as the field $\phi$ reaches its critical
value $\phi_c = M/g$.

The amplitude of  adiabatic density perturbations produced in this
theory
can be estimated by standard methods \cite{MyBook} and is given by
\begin{equation}\label{E24}
\frac{\delta\rho}{\rho} = \frac{16\sqrt {6 \pi}}{5}\
     \ \frac{V^{3/2}}{ {\partial V \over \partial\phi}} ~ = ~
 \frac{16\sqrt {6 \pi}\left({M^4\over 4\lambda} + {m^2\phi^2\over
2}\right)^{3/2}}{5 {m^2\phi}}\ .
\end{equation}
In the case $m^2 \ll H^2$ the scalar field $\phi$ does not change
substantially during the last 60 $e$-foldings (i.e. during the
interval $\Delta t \sim 60 H^{-1}$). In this case the amplitude of
density perturbations practically does not depend on scale, and is
given by
\begin{equation}\label{hybrid2}
{\delta\rho\over \rho} \sim  {2\sqrt{6\pi} g M^5\over 5\lambda
\sqrt{\lambda}  m^2}\ .
\end{equation}
The definition of  ${\delta\rho\over \rho}$ used in   \cite{MyBook}
corresponds to COBE data for  ${\delta\rho\over \rho} \sim 5\cdot
10^{-5}$. Dividing it by (\ref{2a}) with an account taken of
(\ref{hybrid2}) gives $M^3 \ll 5\cdot 10^{-5} \lambda g^{-1} m$. This means
that the `waterfall conditions'
$M^3 \ll    \lambda   \ m $ and $M^3 \ll  {\sqrt \lambda\, g  }
\ m   $ automatically follow from the conditions $m^2\ll H^2$
and ${\delta\rho\over \rho} \sim 5\cdot
10^{-5}$, unless the coupling constants $\lambda$ and $g$ are
extremely small. Therefore the waterfall regime is realized in this
model for a wide variety of  values of parameters $m, M, \lambda$ and
$g$ which lead to density perturbations $\sim 5\cdot 10^{-5}$.

To give a particular example, let us take $g^2 \sim \lambda
\sim10^{-1}$,
$m \sim 10^2$ GeV (electroweak scale). In this case all conditions
mentioned above are satisfied and ${\delta\rho\over \rho} \sim 5\cdot
10^{-5}$ for $M \sim  1.3 \cdot10^{11}$ GeV. In particular, we have
verified, by solving equations of motion for the fields $\phi$ and
$\sigma$ numerically, that inflation in this model ends up within the
time $\Delta t \ll H^{-1}$ after the field $\phi$ reaches its
critical value $\phi_c = M/g$. The value of the Hubble parameter
at the end of inflation is given by $H \sim 7\cdot 10^3$ GeV.  The
smallness of the Hubble constant at the end of inflation
makes it possible, in particular, to have a consistent scenario for
Hybrid in
inflationary cosmology even if the axion mass is much smaller than
$10^{-5}$
eV \cite{Hybrid}. This model has some other distinctive
features. For example, the spectrum of perturbations generated in
this model may look as a power-law spectrum rapidly decreasing at
large wavelength $l$ \cite{LythLiddle}.

Indeed, at the last stages of inflation (for ${M^4\over 4\lambda} \gg
{m^2\phi^2\over 2}$) the field $\phi$ behaves as
\begin{equation}\label{xx}
\phi = \phi_c \cdot \exp{\Bigl(-{m^2(t-t_c)\over 3 H}\Bigr)}\ ,
\end{equation}
whereas the scale factor of the Universe grows exponentially, $a \sim
e^{Ht}$.  This leads to the following relation between  the
wavelength of perturbations $l$ and the value of the scalar field
$\phi$ at the moment when these perturbations were generated: $\phi
\sim \phi_c \left({l\over l_c}\right)^{m^2/3H^2}$. In this case
\begin{equation}\label{xxx}
\frac{\delta\rho}{\rho}  =    {2\sqrt{6\pi} g M^5\over 5\lambda
\sqrt{\lambda}  m^2} \cdot \left({l\over l_c}\right)^{-{m^2\over
3H^2}}\ ,
\end{equation}
which corresponds to the spectrum index $n = 1 + {2m^2\over 3H^2} = 1
+{\lambda m^2\over \pi M^4}$. Note that this spectrum index is
greater than $1$, which is a very unusual feature. For the values of
$m, M, \lambda$ and $g$ considered above, the deviation of $n$ from
$1$ is vanishingly small (which is also  very unusual). However, let
us take, for example, $\lambda = g = 1$, $M = 10^{15}$ GeV (grand
unification scale), and $m =   5\cdot 10^{10}$ GeV. In this case the
amplitude of perturbations at the end of inflation ($\phi = \phi_c$)
is equal to $4\cdot 10^{-4}$,  $n \sim 1.1$, and the amplitude of the
density perturbations drops to the desirable level  ${\delta\rho\over
\rho} \sim 5\cdot 10^{-5}$ on the galaxy scale ($l_g \sim l_c\cdot
e^{50}$). One may easily obtain models with even much larger $n$,
but this may be undesirable, since it may lead to  formation of many
small primordial black holes \cite{Polnarev}.

Note, that the decrease of  ${\delta\rho\over \rho}$ at large $l$ is
not unlimited.  At ${m^2\phi^2\over 2} > {M^4\over 4\lambda}$ the
spectrum begins growing again. Thus, the spectrum has a minimum on a
certain scale, corresponding to the minimum of expression
(\ref{E24}). This complicated shape of the spectrum appears in a very
natural way, without any need to design artificially bent potentials.

As we have seen, coupling constants in our model can be reasonably
large, and the range of possible values of masses $m$ and $M$ is
extremely wide. Thus, our model is very versatile. One should  make
sure, however, that the small effective mass of the
scalar field $\phi$ does not acquire large radiative corrections near
$\phi = \phi_c$. Hopefully this can be done in supersymmetric
theories with flat directions of the effective potential.

\

\section { Reheating after inflation}

The theory of reheating of the Universe after inflation is the
most important application of the quantum theory of
particle creation, since almost all matter constituting the
Universe at the subsequent radiation-dominated stage was
created during this process \cite{MyBook}. At the stage of
inflation all
energy was
concentrated in a classical slowly moving inflaton field $\phi$. Soon
after the
end of inflation this field began to oscillate near the minimum of
its
effective potential. Gradually it produced many elementary particles,
they
interacted with each other and came to a state of thermal equilibrium
with some
temperature $T_r$, which was called the reheating temperature.

An elementary theory  of reheating was first
developed in  \cite{1} for the new inflationary scenario.
Independently a theory of reheating in the $R^2$ inflation was constructed in
\cite{st81}.  Various
aspects of this
theory   were further elaborated by many authors, see e.g.
\cite{Dolg,Brand}.
 Still, a general scenario of reheating was
absent. In particular, reheating in the chaotic inflation theory
remained
almost unexplored.
The present section contains results obtained recently in our work with Kofman
and Starobinsky \cite{KLSREH}. A more detailed presentation of our results is
to be published in
  \cite{REH}.  We have found that the process of
reheating typically
consists of  three different stages.  At the first stage, which cannot be
described by the elementary theory of reheating,  the classical coherently
oscillating
inflaton field $\phi$ decays into massive bosons (in particular, into
$\phi$-particles) due to parametric resonance.  In many models the resonance is
very broad, and the process occurs extremely
rapidly (explosively).  Because of the Pauli exclusion principle, there is no
explosive  creation of fermions.
To distinguish this stage from the stage of  particle decay and thermalization,
we will call it {\it pre-heating}. Bosons produced at that stage are far away
from thermal equilibrium and typically have enormously large occupation
numbers. The second stage  is the decay of previously produced  particles. This
stage typically can be described by methods developed in   \cite{1}. However,
these methods should be applied not  to the decay of the original homogeneous
inflaton field, but to the decay of particles and fields produced at the  stage
of explosive reheating. This considerably changes many features of  the
process, including the final value of the reheating temperature.  The
third stage is the stage of thermalization, which can be described by
standard
methods, see e.g. \cite{MyBook,1}; we will not consider it here.
Sometimes this stage
may occur simultaneously with the second one. In our
investigation we
have used the formalism of the
 time-dependent Bogoliubov transformations   to find the density of
created particles, $n_{\vec k}(t)$.
A detailed
description of this theory will be given in \cite{REH}; here we will
outline
our main conclusions using a simple semiclassical
approach.

We should note that recently two other groups of authors have also studied
related problems, see \cite{Shtanov,deVega}. The corresponding papers are very
useful and contain many interesting results, but some additional work is to be
done in order to apply these results to the theory of reheating after
inflation. For example, ref. \cite{Shtanov} was devoted to study of the stage
of  a narrow resonance without an account taken of backreaction of created
particles. As we already mentioned, the first stages of reheating typically
occur in the regime of  a  broad resonance. On the other hand,  as it was shown
above,  the stage of  a broad resonance ends because of the backreaction of
created particles, and at the  stage of a narrow resonance the effects of
backreaction are also  very important. Backreaction of created particles was
investigated in a very detailed way in ref. \cite{deVega}. However, their
investigation was performed neglecting expansion of the Universe. In the theory
${m^2\over 2}\phi^2  +{\lambda\over 4} \phi^4$ with $m^2 > 0$ the authors
studied the  strong coupling regime, $\lambda \gg 1$, whereas in the chaotic
inflation scenario which we are  investigating now $\lambda \sim 10^{-13}$.
Therefore in what follows we will concentrate on the results obtained in
\cite{KLSREH}.

  We will consider a simple chaotic inflation scenario describing the
classical inflaton scalar field
$\phi$ with the effective potential   $V(\phi) =  \pm {1\over2}
m_\phi^2 \phi^2+{\lambda\over 4}\phi^4$. Minus sign corresponds to
spontaneous symmetry breaking $\phi \to \phi +\sigma$ with generation of a
classical scalar field $\sigma = {m_\phi \over\sqrt\lambda}$.  The field $\phi$
after inflation may decay
into bosons $\chi$ and fermions $\psi$ due to the interaction  terms $- {
1\over2} g^2 \phi^2 \chi^2$ and
 $- h \bar \psi \psi \phi$. Here $\lambda$, $ g$ and  $h$ are
small coupling constants.  In case of  spontaneous symmetry breaking, the term
$- {
1\over2} g^2 \phi^2 \chi^2$ gives rise to the   term  $- g^2 \sigma\phi
\chi^2$. We will assume for simplicity that the bare masses
of the fields $\chi$ and $\psi$ are very small, so that one can write  $ m_\chi
(\phi) =
  g \phi$,  $m_{\psi}(\phi) =  |h\phi|$.

Let us briefly recall the elementary theory of reheating
\cite{MyBook}.  At
$\phi > M_p$, we have a stage of inflation.  This stage is supported
by the
friction-like term $3H\dot\phi$ in the equation of motion for the scalar
field. Here $H\equiv \dot a/a$ is the Hubble parameter, $a(t)$ is the
scale factor of the Universe.
However, with a decrease of the field $\phi$ this term becomes less
and less important, and inflation ends at $\phi {\
\lower-1.2pt\vbox{\hbox{\rlap{$<$}\lower5pt\vbox{\hbox{$\sim$}}}}\ }M_p/2$.
After that the
field $\phi$  begins  oscillating near the minimum of
$V(\phi)$. The amplitude of the oscillations  gradually
decreases because of expansion of the
Universe, and also because of the energy transfer to particles
created by the
oscillating field.  Elementary
theory of reheating is based on the
 assumption that  the
classical oscillating scalar field $\phi (t)$ can be represented as a
collection of scalar particles at rest. Then the rate of decrease of
the energy of oscillations
coincides with the decay rate  of  $\phi$-particles. The
rates of
the processes $\phi \to \chi\chi$  and $\phi \to  \psi\psi$ (for  $m_\phi \gg
2m_\chi, 2m_\psi$) are given
by
 \begin{equation}\label{7}
  \Gamma ( \phi \to \chi \chi) =  { g^4 \sigma^2\over 8
\pi m_{\phi}}\  , \ \ \ \ \
\Gamma( \phi \to \psi \psi )  =  { h^2 m_{\phi}\over 8 \pi}\ .
 \end{equation}
Reheating
completes when the rate of expansion of the Universe given   by the Hubble
constant $H=\sqrt{8\pi \rho\over 3 M^2_p} \sim  t^{-1}$ becomes smaller than
the total decay rate $\Gamma =  \Gamma (\phi \to \chi \chi) + \Gamma
(\phi \to
\psi \psi )$. The reheating temperature can be estimated by
$T_r \simeq 0.1\, \sqrt{\Gamma M_p}$\,.

It is interesting to note that in accordance with the elementary theory of
reheating the amplitude squared of the oscillating scalar field decays
exponentially, as $e^{-\Gamma t}$. Phenomenologically, this can be described by
adding the term $\Gamma\dot\phi$ to the equation of motion of the scalar field.
Unfortunately, many authors took this prescription too seriously and
investigated the possibility that the term $\Gamma\dot\phi$, just like the term
$3H\dot\phi$, can support inflation. We should emphasize \cite{KLSREH}, that
adding the term $\Gamma\dot\phi$ to the equation of motion is justified only at
the stage of oscillations (i.e. after the end of inflation), and only for the
description of the {\it amplitude of oscillations} of the scalar field, rather
than for the description of the scalar field itself.  Moreover, even at the
stage of oscillations this description becomes incorrect as soon as the
resonance effects become important.

As we already mentioned, elementary theory of reheating can provide a
qualitatively correct
description of particle decay at the last stages of reheating.
Moreover, this theory  is always applicable  if the inflaton field
can decay  into fermions only, with a small coupling constant $h^2 \ll
m_{\phi}/M_p$.
 However,
typically this theory is inapplicable to the description of the first stages of
reheating, which makes the whole process quite different. In what follows we
will develop the theory of the first stages of reheating. We will begin with
the theory of a massive scalar field $\phi$ decaying into particles $\chi$,
then we consider the theory  ${\lambda\over 4} \phi^4$,
and finally we will discuss reheating in the theories with spontaneous symmetry
breaking.

We begin with the investigation of the simplest  inflationary
model with the effective potential
${m^2_\phi\over 2}\phi^2$.
Suppose that this field  only interacts
with a light  scalar field $\chi$
 ($m_{\chi} \ll m_{\phi}$) due to the
 term $-{ 1\over2} g^2 \phi^2  \chi^2$.
The equation for quantum fluctuations of the field $\chi$
with the physical momentum $\vec k/a(t)$ has the following form:
\begin{equation}\label{M}
\ddot \chi_k   + 3H \dot \chi_k +  \left({k^2\over a^2(t)}
+ g^2 \Phi^2\, \sin^2(m_{\phi}t) \right) \chi_k = 0 \ ,
\end{equation}
where $k = \sqrt {\vec k^2}$, and $\Phi$ stands for the amplitude of
oscillations of the field $\phi$. As we shall see, the
main contribution to $\chi$-particle
production is given by excitations of the field $\chi$ with
 $k/a \gg m_\phi$, which is much
greater than $H$ at the stage
of oscillations. Therefore, in the first approximation we may neglect
the expansion of the Universe,  taking $a(t)$ as a constant and omitting
the term $3H \dot \chi_k$ in (\ref{M}). Then the equation (\ref{M})
describes an oscillator with a variable
frequency $\Omega_k^2(t)=
 k^2a^{-2} + g^2\Phi^2\, \sin^2(m_{\phi}t) $.
Particle production occurs due to a
nonadiabatic change of this frequency. Equation (\ref{M}) can be
reduced to the well-known  Mathieu equation:
\begin{equation}\label{M1}
\chi_k''   +   \left(A(k) - 2q \cos 2z \right) \chi_k = 0 \ ,
\end{equation}
where   $A(k)
= {k^2 \over m_\phi^2 a^2}+2q$, $q = {g^2\Phi^2\over
4m_\phi^2} $, $z
= m_{\phi}t$, prime denotes differentiation with respect to $z$.
An important property of solutions of the equation (\ref{M1}) is the
existence of an exponential instability $\chi_k \propto \exp
(\mu_k^{(n)}z)$ within the set of resonance bands  of frequencies
$\Delta k^{(n)}$ labeled by an integer index $n$.
This instability corresponds to exponential growth of occupation
numbers of quantum fluctuations
$n_{\vec k}(t) \propto \exp (2\mu_k^{(n)} m_{\phi} t)$
  that may be interpreted as particle
production.   As one
can show,   near the line $A = 2q$ there are regions in the first,
the second and the  higher instability bands
where the unstable modes grow extremely
rapidly, with $\mu_k \sim 0.2$. We will show analytically in
\cite{REH} that  for $q \gg 1$
 typically
 $\mu_k \sim {\ln 3\over 2\pi}
\approx 0.175$ in the instability bands along the line $A = 2q$,
but its maximal value is ${\ln(1+\sqrt{2}) \over \pi} \approx 0.28$.
Creation of
particles in the regime of a broad  resonance ($q > 1$) with $2\pi \mu_k =
O(1)$ is very different from that in  the usually
considered case of a narrow resonance ($ q \ll 1$),
 where $2\pi \mu_k \ll 1$.
 In particular, it
proceeds during a tiny part of each oscillation of the field $\phi$
when $1-\cos z \sim q^{-1}$ and the induced effective mass of the
field $\chi$ (which is
determined by the condition $m^2_{\chi}= g^2\Phi^2/2$) is less than
$m_{\phi}$.
 As a result, the number of
particles grows exponentially within just a few oscillations of the
field
$\phi$. This leads to an extremely rapid  (explosive)  decay of the
classical
scalar field $\phi$.
This regime occurs only
if  $q {\ \lower-1.2pt\vbox{\hbox{\rlap{$>$}\lower5pt\vbox{\hbox{$\sim$}}}}\ }
\pi^{-1}$, i.e. for $g\Phi {\
\lower-1.2pt\vbox{\hbox{\rlap{$>$}\lower5pt\vbox{\hbox{$\sim$}}}}\ }
m_\phi$, so that $m_\phi \ll gM_p$ is the necessary condition for it.
One can show that a typical energy $E$ of a particle produced at this stage is
determined by
 equation $A-2q \sim \sqrt{q}$, and is given by
 $E  \sim  \sqrt{g m_\phi M_p}$ \cite{REH}.

Creation of $\chi$-particles leads to the two main effects:
transfer of the energy from the homogeneous field $\phi (t)$ to these
particles and generation of the contribution to the effective mass of
the $\phi$ field:  $m^2_{\phi ,eff}=m^2_{\phi}+g^2\langle\chi^2
\rangle_{ren}$.
The last term in the latter expression
quickly becomes larger than
$m^2_{\phi}$. One should take
both these effects into account when calculating backreaction of
created particles on the process.
As a result, the stage of the broad resonance creation ends up within
the short time
$t\sim m_{\phi}^{-1} \ln (m_{\phi}/g^5M_p)$,
when $\Phi^2 \sim
\langle\chi^2\rangle$ and  $q = {g^2\Phi^2\over
4m_{\phi ,eff}^2}$
becomes smaller than $1$.
At this time the energy density of produced particles
$\sim E^2 \langle\chi^2\rangle \sim g m_\phi M_p \Phi^2$ is of the same
order as the original energy density
$\sim {m_\phi^2} M_p^2$ of the scalar field
$\phi$ at the end of inflation. This gives the amplitude of
oscillations at the end of the stage of the broad resonance particle creation:
$\Phi^2 \sim \langle\chi^2\rangle \sim
g^{-1} m_\phi M_p \ll M_p^2$.
Since $E\gg m_{\phi}$, the effective equation of state of the whole
system becomes $p\approx \varepsilon /3$. Thus, explosive creation
practically eliminates a prolonged intermediate matter-dominated stage
after the end of inflation which was thought to be characteristic
to many inflationary models.
However, this does not mean that the process of reheating has been completed.
Instead of $\chi$-particles in the thermal equilibrium with
 a typical energy
 $E \sim T \sim (mM_p)^{1/2}$, one has particles with a much
smaller energy $\sim  (g m_\phi M_p)^{1/2}$,
but with extremely large
mean occupation numbers  $n_k \sim g^{-2} \gg 1$.

After that the Universe expands as $a(t)\propto \sqrt t$, and
the scalar field $\phi$ continues its decay in the regime of the narrow
resonance creation $q\approx {\Phi^2\over 4 \langle\chi^2\rangle}
\ll 1$. As a result,
$\phi$ decreases rather slowly, $\phi \propto t^{-3/4}$.
This regime is very important  because  it makes the energy of the
$\phi$ field much smaller than that of the $\chi$-particles.
One can show that the decay finally stops when the amplitude of
oscillations $\Phi$ becomes smaller than $g^{-1} m_\phi$ \cite{REH}.
This happens at the moment $t\sim  m_{\phi}^{-1}  (gM_p/m_{\phi})^{1/3}$
(in the case   $m < g^7 M_p$ decay ends somewhat later,
in the perturbative regime).
The physical reason why the decay  stops is rather general: decay of
the particles $\phi$ in our model occurs due to its interaction with another
$\phi$-particle (interaction term is quadratic in $\phi$ and in $\chi$). When
the  field $\phi$ (or the number of $\phi$-particles) becomes
small, this process
is inefficient.  The scalar field can decay completely only if a
single  scalar $\phi$-particle can decay into other  particles, due
to the processes
$\phi \to \chi \chi$ or $\phi \to \psi \psi$, see eq. (\ref{7}). If
 there is
 no spontaneous symmetry breaking and no interactions with fermions
in our model, such  processes are impossible.

At later stages the energy of oscillations of the inflaton field
decreases as $a^{-3}(t)$, i.e. more slowly than the decrease of
energy of hot ultrarelativistic matter $\propto a^{-4}(t)$. Therefore, the
relative contribution of the field $\phi(t)$ to the total energy density
of the Universe
rapidly grows. This   gives rise to an unexpected possibility that
the inflaton field  by
itself, or other scalar fields can be
cold dark matter candidates, {\it even if they strongly interact with each
other}. However, this possibility requires
a certain degree of fine tuning; a more immediate application of our result is
that it allows one to rule out a wide class of inflationary models which do not
contain interaction terms of the type of $g^2\sigma\phi\chi^2$ or
$h\phi\bar\psi\psi$.

 So far we have not considered  the term ${\lambda \over 4} \phi^4$
in the effective potential. Meanwhile this term leads to production
of $\phi$-particles, which in some cases appears to be the leading
effect.
 Let us  study the $\phi$-particle production in the theory
with $V(\phi)  =
{m^2_{\phi}\over 2} \phi^2 + {\lambda\over 4}\phi^4$ with $m^2_{\phi}
\ll \lambda  M_p^2$. In this case the effective potential
of the field $\phi$ soon after the end of inflation at
$\phi \sim M_p$ is dominated by the term
${\lambda\over 4} \phi^4$.  Oscillations  of the field $\phi$ in this
theory
are not sinusoidal, they are
given by elliptic functions, but with a good accuracy one can write
$\phi(t)
\sim \Phi \sin (c\sqrt \lambda \int \Phi dt)$, where
$c={\Gamma^2(3/4)\over \sqrt \pi} \approx 0.85$.   The Universe at
that time expands as at the
radiation-dominated stage: $a(t)\propto \sqrt t$. If one neglects
the feedback of created $\phi$-particles on the homogeneous field
$\phi (t)$, then its amplitude $\Phi (t) \propto a^{-1}(t)$, so that $a\Phi
=const$.
Using a conformal time $\eta$, exact equation for quantum fluctuations
 $\delta \phi$
 of the field $\phi$ can be reduced to the Lame equation. The results remain
essentially the same if we use an  approximate equation
\begin{equation}\label{lam1}
{d^2(\delta\phi_k)\over d\eta^2}   +   {\Bigl[{k^2} +
3\lambda a^2\Phi^2\, \sin^2 (c\sqrt\lambda a\Phi \eta)\Bigr]}
\delta\phi_k = 0 \ ,~~~\eta =\int {dt\over a(t)}={2t\over a(t)}\, ,
\end{equation}
which leads to the Mathieu equation with $A =
{k^2\over c^2\lambda a^2\Phi^2} +
{3\over 2c^2} \approx  {k^2\over c^2\lambda a^2\Phi^2} + 2.08$, and
$q = {3\over 4c^2} \approx 1.04$.   Looking at the instability chart, we see
that the
resonance occurs in the second band, for $k^2 \sim 3\lambda a^2\Phi^2$. The
maximal value of the coefficient $\mu_k$ in   this band for $q \sim 1$
approximately equals to $0.07$. As long as the backreaction of created
particles is small, expansion of the Universe does not shift fluctuations away
from the resonance band, and the
number of produced particles grows  as  $\exp (2c\mu_k\sqrt\lambda a\Phi \eta)
\sim
\exp ({\sqrt\lambda\Phi t\over 5})$.

After the time interval $\sim M_p^{-1}\lambda^{-1/2}|\ln \lambda|$,
  backreaction of created particles  becomes significant. The growth of the
fluctuations
$\langle\phi^2\rangle$ gives rise to a   contribution
$3\lambda \langle\phi^2\rangle$ to the effective mass squared of the field
$\phi$, both in the equation for $\phi (t)$ and in Eq. (\ref{lam1}) for
inhomogeneous modes.
The stage of explosive reheating ends when $\langle\phi^2\rangle$ becomes
greater than $\Phi^2$. After that, $\Phi^2 \ll
\langle\phi^2\rangle$ and
the effective frequency of oscillations is determined by the
term $\sqrt{3\lambda \langle\phi^2\rangle}$.
The corresponding process is
 described by Eq. (4) with  $A(k) = 1 + 2q + {k^2
\over
3\lambda a^2\langle\phi^2\rangle}$, $q = {\Phi^2\over 4
\langle\phi^2\rangle}$. In this regime $q \ll 1$, and particle creation occurs
in the  narrow resonance regime in the second band with $A \approx 4$.  Decay
of the field in this regime is extremely slow: the amplitude $\Phi$ decreases
only by a factor
 $t^{1/12}$ faster that  it would decrease without any decay, due to the
expansion of the Universe only, i.e., $\Phi \propto t^{-7/12}$ \cite{REH}.
Reheating stops altogether when the presence of non-zero mass
$m_{\phi}$ though still small as compared to $\sqrt{3\lambda
\langle\phi^2\rangle}$
appears enough for the expansion of the Universe to drive
a mode away
from the narrow resonance. It happens when the amplitude $\Phi$ drops
up to a value $\sim m_{\phi}/\sqrt \lambda$.

In addition to this process, the field $\phi$ may decay  to
$\chi$-particles.
This is the leading process for    $g^2\gg \lambda$.
The equation for $\chi_k$ quanta has  the same form as eq.
(\ref{lam1})
with the obvious change $\lambda \to g^2/3$.
Initially  parametric resonance is broad. The values of the parameter
$\mu_k$
along the line $A = 2q$ do not change monotonically, but typically
for $q \gg
1$ they are 3 to 4 times greater than the parameter $\mu_k$ for the
decay of
the field $\phi$ into its own quanta. Therefore, this pre-heating
process is very
efficient. It ends at the moment $t\sim M_p^{-1}\lambda^{-1/2}
\ln (\lambda /g^{10})$ when $\Phi^2 \sim \langle \chi^2
\rangle \sim g^{-1}\sqrt \lambda M_p^2$. The typical energy of created
$\chi$-particles is $E \sim (g^2\lambda)^{1/4}M_p$. The following
evolution is essentially the same as that described above for the case of a
massive scalar field decaying into $\chi$-particles.

Finally, let us consider the case with symmetry breaking.  In the
beginning, when the amplitude of oscillations is much greater than
$\sigma$, the theory of  decay of the inflaton field is the same as in the case
considered above. The most important part of  pre-heating occurs at this stage.
When the amplitude of the oscillations becomes smaller than
$m_\phi/\sqrt\lambda$ and the field begins oscillating near the minimum of the
effective potential at $\phi = \sigma$, particle production due to
the narrow parametric
resonance typically becomes  very weak.
The main reason for this is related to the backreaction of
particles created at the
preceding stage of pre-heating on the rate of expansion of the Universe and on
 the shape of the effective potential \cite{REH}. However, importance of
spontaneous symmetry
breaking for the theory of reheating should not be underestimated, since it
gives rise to the interaction term   $g^2\sigma\phi\chi^2$ which is linear in
$\phi$. Such terms are necessary for a complete decay of the inflaton field in
accordance with the perturbation theory (\ref{7}).

Let us briefly summarize our results:

1. In many models where decay of the inflaton field can occur in the purely
bosonic sector the first stages of reheating occur due to parametric resonance.
This process (pre-heating) is extremely efficient even if the corresponding
coupling constants are very small.
However, there is no explosive reheating in the models where decay of the
inflaton field is necessarily accompanied by fermion production.

 2. The stage of explosive reheating due to a broad resonance typically is very
short. Later the resonance becomes narrow, and
finally the stage of pre-heating finishes altogether. Interactions of
particles produced at this stage, their decay into other particles and
subsequent thermalization typically
require  much more time that the stage of pre-heating, since these processes
are suppressed by the small
values of coupling constants.

3. The last stages of reheating typically
  can be described by the elementary theory of reheating \cite{1}. However,
this  theory  should be applied not to the decay of the
original large and homogeneous oscillating inflaton field, but to the decay  of
particles produced at the stage of pre-heating, as well as to the decay of
small remnants of the  classical   inflaton field. This makes a lot
of difference, since typically coupling constants of interaction of the
inflaton field with matter are extremely small, whereas coupling constants
involved in the decay of  other bosons  can be much greater. As a result,
the reheating temperature can be much higher than the typical temperature $T_r
{\ \lower-1.2pt\vbox{\hbox{\rlap{$<$}\lower5pt\vbox{\hbox{$\sim$}}}}\ } 10^9$
GeV
which could be obtained
neglecting the stage of parametric resonance \cite{REH}.
On the other hand, such processes as baryon creation after inflation occur best
of all outside  the state of thermal equilibrium. Therefore, the stage
of   pre-heating  may play an extremely important role in our cosmological
scenario.  Another  consequence of the resonance effects is an almost
instantaneous change of equation of state from the vacuum-like one to the
equation of state of relativistic matter.

4. Reheating never completes in the theories where a single $\phi$-particle
cannot decay into other particles. This implies that reheating completes only
if the theory  contains interaction terms like $\phi\sigma\chi^2$ of
$\phi\bar\psi\psi$. In most cases the theories where reheating never completes
contradict observational data.  On the other hand, this result suggests an
interesting possibility that  the classical scalar fields  (maybe even the
inflaton field itself) may be responsible for the
dark matter of the Universe even if they strongly interact with other matter
fields.

\

\section { Topological defects as seeds for eternal inflation}

 One of the main purposes of inflationary cosmology was to solve the primordial
monopole problem.  The solution was very simple: inflation exponentially
increases the distance between monopoles and makes their density negligibly
small.
It was assumed that inflation neither can  change internal properties of
monopoles nor can it lead to their copious production. For example, in the
first version of the new inflation scenario based on the $SU(5)$
Coleman-Weinberg theory \cite{b16} the Hubble constant during inflation was of
the order $10^{10}$ GeV, which is five orders of magnitude smaller than the
mass of the $X$-boson $M_X \sim 10^{15}$ GeV. The size of a monopole given by
$\sim M_X^{-1}$   is five orders of magnitude smaller than the  curvature of
the Universe given by the size of the horizon $H^{-1}$. It seemed obvious that
such monopoles simply could not know that the Universe is curved.

This conclusion finds an independent confirmation in the calculation of the
probability of spontaneous creation of monopoles during inflation. According to
\cite{GuthVil}, this probability is suppressed by   a factor of $\exp (-2\pi
m/H)$, where $m$ is the monopole mass. In the model discussed above this
factor is given by $\sim 10^{-10^6}$, which is negligibly small.

Despite all these considerations, in the present section   (see also \cite{LL})
we will show  that in many theories, including the $SU(5)$ Coleman-Weinberg
theory, monopoles, as well as other topological defects,  do expand
exponentially and can be copiously produced during inflation. The main reason
is very simple. In the center of a  monopole the scalar field $\phi$ always
vanishes. This means that monopoles always stay on the top of the effective
potential at $\phi = 0$. When inflation begins, it makes the field $\phi$
almost homogeneous and very close to zero near the center of the monopole. This
provides excellent conditions for inflation inside the monopole. If inflation
is supported by the vacuum energy $V(0)$, the conditions for inflation  remain
satisfied inside the monopoles even after inflation finishes outside them. An
outside observer will consider such monopoles as magnetically charged
Reissner-Nordstr\"om black holes. A typical distance between such objects will
be exponentially large, so they will not cause any cosmological problems.
However, each such magnetically charged black hole will have an inflationary
universe inside it.  And, as we are going to argue, each such inflationary
universe will contain many other inflating monopoles, which after inflation
will look like black holes with inflationary heart. We will call such
configurations {\it fractal monopoles}, or, more generally, {\it fractal
topological defects}.

   To explain the basic idea in a more detailed way, we will consider first the
simple model with the Lagrangian
\begin{equation}\label{1u}
L= {1\over 2} (\partial _\mu \phi )^2 - {\lambda\over 4} \Bigl(\phi^2
  - {m^2\over \lambda} \Bigr)^2  \ ,
\end{equation}
where  $\phi$ is a real scalar field.
Symmetry breaking in this model leads to formation of domains with $\phi = \pm
\eta$, where $\eta = {m\over \sqrt \lambda}$. These domains are divided by
domain walls   (kinks) which interpolate between
the two minima. Let us for a moment neglect gravitational effects. In this case
one can easily obtain a solution for a static domain wall  in the $yz$ plane:
\begin{equation}\label{2u}
\phi = \eta\ {\tanh} \Bigl( \sqrt{\lambda\over 2}\ \eta x  \Bigr) \ .
\end{equation}

Now let us see whether our neglect of gravitational effects is reasonable. The
potential energy density in the center of the kink (\ref{2u}) at $x = 0$ is
equal to ${\lambda\over 4}  \eta ^4$, the gradient energy is also equal to
${\lambda\over 4}  \eta ^4$. This energy density remains almost constant  at
$|x| \ll m^{-1} \equiv {1\over \sqrt \lambda \eta}$, and then it rapidly
decreases.  Gravitational effects can be neglected if the Schwarzschild radius
$r_g = {2M}$ corresponding to the distribution of matter with energy density
$\rho = {\lambda\over 2}  \eta ^4$ and radius $R \sim m^{-1}$ is much smaller
than $R$. Here $M = {4\pi\over 3} \rho R^3$. This condition suggests that
gravitational effects can be neglected for $\eta \, \ll \, {3\over 2\pi}$. In
the opposite case,
\begin{equation}\label{3u}
\eta \, {\ \lower-1.2pt\vbox{\hbox{\rlap{$>$}\lower5pt\vbox{\hbox{$\sim$}}}}\ }
\, {3\over 2\pi}   \ ,
\end{equation}
gravitational effects can be very important.

This result remains true for other topological defects as well. For example,
recently it was shown that monopoles in the theory with the scale of
spontaneous symmetry breaking $\eta {\
\lower-1.2pt\vbox{\hbox{\rlap{$>$}\lower5pt\vbox{\hbox{$\sim$}}}}\ }$ form
Reissner-Nordstr\"om black holes \cite{Lee}. It remained unnoticed, however,
that the same condition, $\eta {\
\lower-1.2pt\vbox{\hbox{\rlap{$>$}\lower5pt\vbox{\hbox{$\sim$}}}}\ } 1$, is
simultaneously a condition of inflation  at $\phi \ll \eta$ in the model
(\ref{1u}).

Indeed, inflation occurs at $\phi \ll \eta$ in the model (\ref{1u}) if the
curvature of the effective potential $V(\phi)$ at $\phi \ll \eta$ is much
smaller than $3 H^2$, where $H = \sqrt{2\pi\lambda\over 3}\, {\eta^2 }$ is the
Hubble constant supported by the effective potential \cite{MyBook}. This gives
$m^2 \ll {2\pi\lambda \eta^4 }$, which leads to the condition almost exactly
coinciding with (\ref{3}): $\eta \, \gg /\sqrt{2\pi}$.

This coincidence by itself does not mean that domain walls and monopoles in the
theories with $\eta \, \gg \,  1/\sqrt{2\pi}$ will inflate. Indeed, inflation
occurs only if the energy density is dominated by the vacuum energy. As we have
seen, for the wall (\ref{2u}) this was not the case:  gradient energy density
for the solution (\ref{2u}) near $x = 0$  is equal to the potential energy
density. However, this is correct only after inflation and only if
gravitational effects are not taken into account.

At the initial stages of inflation the field $\phi$ is equal to zero. Even if
originally there were any gradients of this field, they rapidly become
exponentially small. Each time $\Delta t = H^{-1}$ new  perturbations with the
 amplitude ${H/\sqrt{2} \pi}$ and the wavelength $\sim H^{-1}$ are produced,
but their gradient energy density $\sim H^4$ is always much smaller than
$V(\phi)$ for $V(\phi) \ll 1$ \cite{LLM}.

Formation of domain walls in this scenario can be explained as follows. Assume
that in the beginning the field $\phi$ inside some domain of initial size
$O(H^{-1})$ was equal to zero. It did not move classically at that time, since
$V'(0) = 0$.  However, within the time $\Delta t = H^{-1}$ fluctuations of the
field $\phi$ were generated, which looked like sinusoidal waves with an
amplitude ${H/\sqrt{2} \pi}$ and wavelength $\sim H^{-1}$. During this time the
original domain grows in size $e$ times, its volume grows $e^3 \sim 20$ times.
Therefore it becomes divided into $20$ domains of a size of the horizon
$H^{-1}$. Evolution of the field inside each of them occurs independently of
the processes in the other domains (no-hair theorem for de Sitter space). In a
half of these domains scalar field will have   average value $\phi \approx  +
{H/ {2} \pi}$, in other domains it will have   average value $\phi \approx  -
{H/ {2} \pi}$. After that, the field $\phi$   begins its classical motion. In
those   domains where the resulting amplitude of the fluctuations is positive,
the field $\phi$ moves towards the minimum of the effective potential at $\phi
= + \eta$; in other domains it moves towards $\phi = - \eta$. These domains
become separated by the domain walls with $\phi = 0$. However, as we already
mentioned, at this stage the gradient energy near these walls is similar to
$H^4 \sim  V^2(\phi) $, which is much smaller than $V(\phi)$ for  $V(\phi) \ll
1$. Thus, domain walls continue expanding exponentially {\it in all
directions}.

Domain wall formation is not completed at this point yet. Indeed, the field
$\phi$ may jump back from the region with $\phi > 0$ to $\phi < 0$. However,
the probability of such jumps is smaller than $1/2$, and it becomes even much
smaller with a further growth of $|\phi|$. What is more important, the typical
wavelength of new fluctuations remains $H^{-1}$, whereas the previously formed
domains with $\phi \sim  \pm {H/ {2} \pi}$ continue growing exponentially.
Therefore the domains with negative $\phi$ produced by the jumps of the field
$\phi$ will appear inside   domains with positive $\phi$, but they will look
like   small islands with $\phi < 0$ inside the sea with $\phi >0$. This means
that the subsequent stages of inflation cannot destroy the originally produced
domain walls; they can only produce new domain walls on a smaller length scale.
These new walls will be formed only in those places where the scalar field is
sufficiently small for the jumps with the change of the sign of the field
$\phi$ to be possible. Therefore the new walls will be created predominantly
near the old ones (where $\phi = 0$), thus forming a fractal domain wall
structure.

\

 Similar effects occur in more complicated models where instead of  a discrete
symmetry $\phi \to - \phi$ we have a continuous symmetry. For example, instead
of the model (\ref{1u}) one can consider a model
\begin{equation}\label{5u}
L= \partial _\mu \phi^*\,  \partial _\mu \phi  - {\lambda} \left(\phi^*\phi
  - {\eta^2\over 2 } \right)^2  \ ,
\end{equation}
where  $\phi$ is a complex scalar field, $\phi = {1\over \sqrt 2}(\phi_1 + i
\phi_2)$.
Spontaneous  breaking of the $U(1)$ symmetry in this theory may produce global
cosmic strings. Each string contains  a line with $\phi = 0$. Outside this line
the absolute value of the field $\phi$ increases and asymptotically approaches
the limiting value  $\sqrt{\phi_1^2 + \phi_2^2} = \eta$. This string will be
topologically stable if the isotopic vector $(\phi_1(x), \phi_2(x))$ rotates by
$2n\pi$ when the point $x$ takes a closed path  around the string.

The next step is to consider a theory with $O(3)$ symmetry,
\begin{equation}\label{6u}
L= {1\over 2}  (\partial _\mu\vec \phi )^2 -
{\lambda \over 4} ( {\vec \phi}^2 - {{\eta ^ 2}} ) ^2
  \ ,
\end{equation}
where $\vec \phi$ is a vector $(\phi_1, \phi_2, \phi_3)$. This theory admits
global monopole solutions. The simplest monopole configuration contains a point
$x = 0$ with $\phi(0) = 0$ surrounded by the scalar field $\vec \phi(x) \propto
\vec x$. Asymptotically this field approaches regime with $\vec \phi^2(x) =
\eta^2$.

The most important feature of strings and monopoles is the existence of the
points where $\phi = 0$. Effective potential has an extremum at $\phi = 0$, and
if the curvature of the effective potential is smaller than $H^2 = {8\pi
V(0)\over 3 }$, space around the points with $\phi = 0$ will expand
exponentially, just as in the domain wall case considered above.

Now we will add gauge fields. We begin with the Higgs model, which is a direct
generalization of the model (\ref{5u}):
\begin{equation}\label{5a}
L= D _\mu \phi^*\,  D _\mu \phi  -  {1\over 4} F _{\mu \nu} F^{\mu \nu}   -
{\lambda} \left(\phi^*\phi
  - {\eta^2 \over 2 }\right)^2  \ .
\end{equation}
Here $D _\mu$ is a covariant derivative of the scalar field, which in this
simple case (Abelian theory) is given by $\partial _\mu - i e A_\mu$. In this
model strings of the scalar field contain  magnetic flux $\Phi = 2\pi/e$. This
flux is localized near the center of the string with $\phi(x) = 0$, for the
reason that the vector field becomes heavy at  large $\phi$, see e.g.
\cite{Kirzhnits}. However, if inflation  takes place inside the string, then
the field $\phi$ becomes vanishingly small not only at the central line with
$\phi(x) = 0$, but even exponentially far away from it. In such a situation the
flux of magnetic field will not be confined near the center of the string. The
thickness of the flux will grow together with the growth of the Universe. Since
the total flux of magnetic field inside the string is conserved, its strength
will decrease exponentially,
and very soon its effect on the  string expansion will become negligibly small.
A deep underlying reason for this behavior is the conformal invariance of
massless vector fields. Energy density of such fields decreases as $a(t)^{-4}$,
where $a(t)$ is a scale factor of the Universe. The final conclusion is that
the vector fields do not prevent inflation of strings. The condition for
inflation to occur inside the string remains the same as before: the curvature
of the effective potential at $\phi = 0$ should be smaller than $H^2 = {8\pi
V(0)\over 3 }$.

The final step is to consider  magnetic monopoles. A simplest example is given
by the $O(3)$ theory
\begin{equation}\label{7u}
L= {1\over 2}  |(D _\mu {\vec \phi} | ^2  - {1\over 4} F^a _{\mu \nu} F^{a \mu
\nu}-
{\lambda \over 4} ( {\vec \phi}^2 - {{\eta ^ 2}}  ) ^2  \ .
\end{equation}
Global monopoles of the theory (\ref{5u}) become magnetic monopoles in the
theory (\ref{5a}). They also have $\phi = 0$ in the center.  Vector fields in
the center of the monopole are massless ($g\phi = 0$). During inflation these
fields exponentially decrease, and therefore they do not affect inflation of
the monopoles.

We should emphasize that even though the field $\phi$ around the monopole
during inflation is very small, its topological charge is well defined,  it
cannot change and it cannot annihilate with the charge of other monopoles as
soon as the radius of the monopole becomes greater than $H^{-1}$.  However, an
opposite process is possible. Just as domain walls can be easily produced by
quantum fluctuations near other inflating domain walls, pairs of monopoles can
be produced in the vicinity of an inflationary monopole. The distance between
these monopoles grow exponentially, but the new monopoles will appear in the
vicinity of each of them. This picture has been confirmed by the results of
computer simulation of this process \cite{LL}.

Note that in the simple models discussed above inflation of monopoles occurs
only if spontaneous symmetry breaking is extremely strong, $\eta {\
\lower-1.2pt\vbox{\hbox{\rlap{$>$}\lower5pt\vbox{\hbox{$\sim$}}}}\ } 1$.
However, this is not a necessary condition. Our arguments remain valid for {\it
all} models where the curvature of the effective potential near $\phi = 0$ is
smaller than the Hubble constant supported by $V(0)$. This condition is
satisfied by all models which were originally proposed for the realization of
the new inflationary universe scenario.
In particular, the monopoles in the $SU(5)$ Coleman-Weinberg theory also should
expand exponentially. However, to make sure that we do not miss something
important, we should resolve the paradox     formulated in the very beginning
of this section: The Hubble constant $H$ during inflation in the $SU(5)$
Coleman-Weinberg theory is much smaller than the mass of the vector field
$M_X$, which is usually related to the size of the monopole. In such a
situation inflation of the interior of the monopole does not seem possible.

This objection is very similar to the argument that the domain wall should not
inflate since its gradient energy density is equal to its potential energy
density. The resolution of the paradox is very similar. The effective mass of
the vector field $M_X  \sim g\eta \sim 10^{15}$ GeV can determine  an effective
size of the monopole only {\it after} inflation.   Effective mass of the vector
field $M_X(\phi) \sim g\phi$ is always equal to zero in the center of the
monopole. Once inflation begins in a domain of a size $O(H^{-1})$ around the
center of the monopole, it expels vectors fields away from the center and does
not allow them to penetrate back as far as inflation continues.

\

As we already discussed, it in many versions of inflationary theory some parts
of the Universe expand without end. Our results reveal a very nontrivial role
which topological defects may play in this process.  For example, in the theory
with breaking of a discrete symmetry $\phi \to - \phi$  the Universe in the
source of its evolution   becomes divided into many thermalized regions divided
by exponentially expanding domain walls unceasingly producing new inflating
walls.
It is important that inflation {\it never stops} near the walls. Some of these
walls could collapse, eating the island of thermalized phase inside them.
However, it does not seem  possible. Due to the no-hair theorem for de Sitter
space, the part of the Universe near the wall (i.e. near $\phi = 0$) lives by
its own laws and continue expanding exponentially all the time.  Thus, the
walls become indestructible  sources of eternal inflation, which produce new
inflating walls, which produce new walls, etc. Far away from the walls, the
field $\phi$ relaxes near $\phi = \pm \eta$.  However, due to  the exponential
expansion of space near the walls, they always remain exponentially thick and
never approach the thin wall solution described by eq. (\ref{2}).

The picture of eternally inflating Universe consisting of  islands of
thermalized phase surrounded by inflating domain walls is very similar to a
picture which appears in old inflation and in the versions of new inflation
where the state $\phi = 0$ is sufficiently stable. However,  in our model the
main reason  for this behavior is purely  topological \cite{LL}. If inflation
in this theory begins at large $\phi$, as in the simplest models of chaotic
inflation, then the Universe will not consist of islands of thermalized phase
surrounded by de Sitter space. On the contrary, it will consist of islands of
inflationary universe surrounded by the thermalized phase \cite{LLM}.
Nevertheless, inflation in this case will go eternally due to the process of
self-reproduction of inflationary domains with large $\phi$ \cite{b19}.

 The results of computer simulations performed in \cite{LL} suggest that in the
theories with continuous symmetry breaking, which allow existence of inflating
strings and monopoles, the global structure of the Universe is similar to the
structure of the Universe in the simplest versions of chaotic inflation. The
Universe will consist of islands of inflationary phase associated with
inflating monopoles (or of lines of inflationary phase associated with
inflating strings) surrounded by thermalized phase. Since   monopoles are
topologically stable, and field $\phi$ is equal to zero in their centers,  they
form indestructible sources of inflation. The structure of space-time near each
such monopole is very complicated; it should be studied by the methods
developed in \cite{Berezin} for description of a bubble of de Sitter space
immersed into vacuum with vanishing energy density.  Depending on initial
conditions, many possible configurations may appear. The simplest one is the
monopole which looks like a small magnetically charged black hole from the
outside, but which contains a part of exponentially expanding space inside it.
This is a wormhole configuration similar to those   studied in
\cite{Markov}--\cite{Tkachev}. The physical interpretation of this
configuration can be given as follows \cite{Berezin,Tkachev}. An external
observer will   see a small magnetically charged Reissner-Nordstr\"om black
hole.  This part of the picture is consistent with the results obtained in
\cite{Lee}. On the other hand,   an observer near  the monopole  will see
himself inside an eternally inflating part of the Universe. These two parts of
the Universe will be connected by a wormhole. Note that the wormhole connecting
  inflationary magnetic monopole to our space cannot evaporate unless the
inflationary universe can loose its magnetic charge.

At the quantum level the situation becomes even more interesting. As we already
mentioned, fluctuations of the field $\phi$ near the center of a monopole are
strong enough to create new regions of space with $\phi = 0$, some of which
will become   monopoles. After a while, the distance between these monopoles
becomes exponentially large, so that they cannot annihilate. This process of
monopole-antimonopole pair creation produces a fractal structure consisting of
monopoles created in the vicinity of other monopoles.  Each of these monopoles
in the process of its further evolution will evolve into   a black hole
containing inflationary universe containing many other monopoles, etc. That is
why we call them {\it fractal monopoles}.

Note that one of the main motivations for the development of inflationary
cosmology was a desire to get rid of primordial monopoles. However, the
original version of the new inflationary universe scenario based on the theory
of high temperature phase transitions did not work particularly well, since it
required rather unnatural initial conditions \cite{MyBook}. Fortunately, the
same potentials which gave rise to new inflation can be used in the context of
the chaotic inflation scenario, which does not require the whole Universe being
in a state of thermal equilibrium with $\phi = 0$. If these initial conditions
can be realized at least in one domain of initial size $O(H^{-1})$, the
Universe enters the stage of eternal inflation. Of course, this possibility is
still rather problematic since in the models used in the new inflationary
universe scenario the size $O(H^{-1})$ is  many orders of magnitude greater
than the Planck length. For example, the probability of  quantum creation of
inflationary universe in the new inflationary universe scenario is suppressed
by an infinitesimally small factor $\exp\Bigl(-{3 \over 8V(0)}\Bigr) \sim
10^{-10^{16}}$ \cite{Creation}. Therefore it is much easier for inflation to
begin and continue eternally in the simplest chaotic inflation models where
inflation is possible even at $H \sim 1$, $V(\phi) \sim 1$ \cite{LLM,b19}.
Still the problem of initial conditions in the models with the potentials used
in the new inflationary scenario does not look that bad if one takes into
account that inflation in these models also goes on  without end
\cite{b52,b62}, and eventually we  become generously rewarded for our initial
problems by the infinite growth of  the total volume of  inflationary domains.
Moreover, as it was shown in \cite{LLM}, the problem  of initial conditions for
the models with the potentials used in the new inflationary scenario can be
easily solved by a preceding stage of chaotic inflation. Now we encounter a new
amazing twist of   this scenario.  Eternal inflation begins if initially we
have at least one inflating monopole. Such monopoles exponentially expand,
create new space suitable for life of our type to appear there, and then these
monopoles effectively expel themselves from the space where inflation ended by
continuous creation of exponentially expanding space around them. The
possibility that monopoles by themselves can solve the monopole problem is so
simple that it certainly deserves further investigation.

\

 \section { Stationary Universe}

Even though eternal inflation due to topological defects looks very
interesting, it is much less general than the eternal inflation which occur
because of quantum fluctuations in many other theories including the simplest
models of chaotic inflation with the potentials ${m^2\over 2}\phi^2$ or
${\lambda\over 4}\phi^4$ \cite{b19}.
In fact, the effect of self-reproduction of inflationary universe occurs almost
in all versions of inflationary cosmology (even though some exceptions are
possible \cite{BL}). This forces us to reconsider many features of the   big
bang cosmology{\bf {\bf }}.

 The first models of inflation were  based on the standard assumption
of the
big bang theory that the Universe was created at a single moment of
time in a
state with the Planck density, and that it was hot and large (much
larger than
the Planck scale $M_p^{-1} =1$) from the very beginning. The  success
of
inflation in solving internal problems of the big bang  theory
apparently
removed the last doubts concerning the  big bang cosmology.  It
remained almost
unnoticed that during the last ten years the inflationary theory  has
broken
the umbilical cord connecting it with the old big bang theory, and
acquired an
independent life of its own. For the practical purposes of
description of the
observable part of our Universe one may still speak about the big bang.
However, if one tries to understand the beginning of the Universe, or
its end,
or its global structure, then some of the notions of the big bang
theory become
 inadequate.

For example, already in the first version of the   chaotic inflation
scenario \cite{b17} there was no need to assume that the whole Universe
appeared from
nothing at a single moment of time associated with the big bang, that
the
Universe was hot from the very beginning and that the inflaton scalar
field
$\phi$ which drives inflation originally occupied the minimum of its
potential
energy. Later it was found that if the Universe contains at least one
inflationary domain of a size of horizon  (`$h$-region') with a
sufficiently
large and homogeneous scalar field $\phi$, then this domain will
permanently
produce new $h$-regions of a similar type. In other words,  instead
of a single
big bang producing a one-bubble Universe, we are speaking now about
inflationary bubbles producing new bubbles, producing new bubbles,
{\it ad
infinitum}. In this sense, inflation is not a short intermediate
stage of
duration $\sim 10^{-35}$ seconds, but a self-regenerating process,
which occurs
in some parts of the Universe even now, and which will continue
without end.

Thus, recent developments of inflationary theory have considerably
modified our
cosmological paradigm \cite{MyBook}.  Now  we must learn  how to
formulate
physical questions in the new context. For example, in a homogeneous
part of
the Universe there is a simple relation between  the density of
matter and
time. However, on a very  large scale the Universe becomes extremely
inhomogeneous. Its density, at the same `cosmic time', varies
anywhere from
zero to the Planck density. Therefore the question about the density
of the
Universe at the time $10^{10}$ years may not have any definite
answer. Instead
of  addressing such questions we should study the distribution of
probability
of finding a part of the Universe with  given properties, and   find
possible
correlations between these properties.

It is extremely complicated to describe an inhomogeneous Universe and
to find
the corresponding probability distribution.  Fortunately, there
exists a
particular
kind of stationarity of the process of the Universe self-reproduction
which
makes things  more regular. Due to the no-hair theorem for de Sitter
space, the
process of production of new inflationary domains occurs
independently of any
processes outside the horizon. This process depends only on the
values of the
fields inside each $h$-region   of  radius $H^{-1}$. Each time  a new
inflationary $h$-region is created during the Universe expansion, the
physical
processes inside this region  will depend only on the properties of
the fields
inside it, but  not  on the `cosmic time'  at which it was created.

In addition to this most profound stationarity,  which we will call {\it local
stationarity } there may also exist some
simple stationary probability distributions which may allow us to
say, for
example,  what  the probability is of finding a given field $\phi$ at
a given
point.  To examine this possibility one should consider the
probability
distribution $P_c(\phi, t|\chi)$, which  describes the probability
of finding
the field $\phi$ at a given point at a time $t$, under the condition
that at
the time $t=0$ the field $\phi$ at this point was equal to $\chi$. The same
function may also describe the probability
that the
scalar field which at  time $t$ was equal to $\phi$, at some earlier
time $t=0$
was equal to $\chi$.

The probability distribution $P_c$ has been studied by many authors,
see e.g.
refs. \cite{b19,b60,Bond}. Our  investigation of this
question has
shown that in all realistic inflationary models the probability
distribution
$P_c(\phi, t|\chi)$ is not stationary \cite{b19}. The reason
is very
simple.   The  probability distribution $P_c$  is  in fact the
probability
distribution per  unit volume in {\it comoving coordinates} (hence
the index
$c$ in  $P_c$), which do  not change during expansion of the
Universe.  By
considering this probability distribution   we neglect the main
source of the
self-reproduction of inflationary domains, which is the exponential
growth of
their volume. Therefore, in addition to   $P_c$, we introduced the
probability
distribution $P_p(\phi, t|\chi)$, which describes the probability to
find a
given field configuration in a unit physical volume \cite{b19}. In
this section  we
will show
that under certain conditions the stationary probability distribution
$P_p(\phi, t|\chi)$ does exist \cite{LLM}. In such situations we will speak
about {\it global stationarity}.

First of all we should remember some details of stochastic approach
to
inflation.
Let us consider the simplest model of chaotic inflation based on the
theory of
a  scalar field $\phi$ minimally coupled to gravity, with the
effective
potential $V(\phi)$. If the classical field $\phi$ is sufficiently
homogeneous
in some domain of the Universe, then its behavior inside this domain
is
governed by the equation $3H\dot\phi = -dV/d\phi$, where $H^2 =
\frac{8V(\phi)\pi}{3 }$.

Investigation of these equations has shown that  in all power-law potentials
$V(\phi)\sim \phi^n$ inflation occurs
at $\phi > \phi_e \sim n/6$. In the theory with an exponential potential
$V(\phi)\sim
e^{\alpha \phi}$ inflation ends only if we bend down the potential at some
point $\phi_e$;
for definiteness we will take $\phi_e = 0$ in this theory.

Inflation stretches all initial inhomogeneities. Therefore, if the
evolution of the Universe were governed solely by classical equations
of
motion, we would end up with an extremely smooth Universe with no
primordial
fluctuations to initiate the growth of galaxies.
Fortunately, new density perturbations are generated during inflation
due to
quantum effects. The wavelengths of all  vacuum
fluctuations of the scalar field $\phi$ grow exponentially in the
expanding
Universe. When the wavelength of any particular fluctuation becomes
greater
than $H^{-1}$, this fluctuation stops oscillating, and its amplitude
freezes at
some nonzero value $\delta\phi (x)$ because of the large friction
term
$3H\dot{\phi}$ in the equation of motion of the field $\phi$\@. The
amplitude
of this fluctuation then remains
almost unchanged for a very long time, whereas its wavelength grows
exponentially. Therefore, the appearance of such a frozen fluctuation
is
equivalent to the appearance of a classical field $\delta\phi (x)$
that does
not vanish after averaging over macroscopic intervals of space and
time.

Because the vacuum contains fluctuations of all wavelengths,
inflation leads to
the creation of more and more perturbations of the classical field
with
wavelengths greater than $H^{-1}$\@. The average amplitude of such
perturbations generated during a time interval $H^{-1}$ (in which the
Universe
expands by a factor of e) is given by
\begin{equation}\label{E23}
|\delta\phi(x)| \approx \frac{H}{2\pi}\ .
\end{equation}
The phases of each wave are random. Therefore,  the sum of all waves
at a given
point fluctuates and experiences Brownian jumps in all directions in
the space
of fields.

The standard way to describe  the stochastic behavior of the inflaton
field during the slow-rolling stage is to coarse-grain it over
$h$-regions
and consider the effective equation of motion of the long-wavelength
field \cite{b60}:
 \begin{equation} \label{m1}
\frac{d \phi}{dt} = -\, \frac{V'(\phi)}{3H(\phi)} +
\frac{H^{3/2}(\phi)}{2\pi}
\, \xi(t) \ .
\end{equation}
Here  $H= \sqrt{8\pi V/3}$\,, \, $\xi(t)$ is the effective white
noise
generated by quantum fluctuations,
which leads to the Brownian motion of the classical field $\phi$\@.

This Langevin equation leads to two stochastic equations for the
probability
distribution $P_c(\phi,t|\chi)$. The first one is called the backward
Kolmogorov equation,
\begin{equation} \label{b60b}
\frac{\partial  P_c(\phi,t|\chi)}{\partial t} =
 \frac{1}{2} \frac{H^{3/2}(\chi)}{2\pi}\, \frac{\partial
}{\partial\chi}
 \left( \frac{H^{3/2}(\chi)}{2\pi} \, \frac{\partial }{\partial\chi}
P_c(\phi,t|\chi)
\right)
 -  \frac{V'(\chi)}{3H(\chi)} \frac{\partial }{\partial\chi}
P_c(\phi,t|\chi)   \ .
\end{equation}
In this equation one considers the value of the field $\phi$ at the
time \, $t$
as a constant,  and finds the time dependence of the probability that
this
value  was reached  during the time \,  $t$ as a result of
diffusion of the
scalar field from different possible initial values $\chi \equiv
\phi(0)$.

The second equation is the adjoint to the first one; it is called the
forward
Kolmogorov equation, or the Fokker-Planck equation \cite{b60},
\begin{equation}\label{E3711}
\frac{\partial P_c(\phi,t|\chi)}{\partial t} =
 \frac{1}{2}  \frac{\partial }{\partial\phi}
 \left( \frac{H^{3/2}(\phi)}{2\pi}\, \frac{\partial }{\partial\phi}
\Bigl(
 \frac{H^{3/2}(\phi)}{2\pi}  P_c(\phi,t|\chi) \Bigr) +
\frac{V'(\phi)}{3H(\phi)} \, P_c(\phi,t|\chi)\right) \ .
\end{equation}

One may try to find a stationary solution of equations (\ref{b60b}),
(\ref{E3711}), assuming that  $\frac{\partial
P_c(\phi,t|\chi)}{\partial t} =
0$.
The simplest stationary solution  (subexponential factors being
omitted) would
be
\begin{equation}\label{E38a} P_c(\phi,t|\chi) \sim
\exp\left({3\over 8 V(\phi)}\right)\cdot \exp\left(-{3\over 8
V(\chi)}\right) \
{}.
\end{equation}
This function is extremely interesting. Indeed, the first term  in
(\ref{E38a})
is equal to the square of the Hartle-Hawking wave function of the
Universe  \cite{HH}, whereas the second one gives the square of the tunneling
wave
function   \cite{Creation}\hskip .1cm !

At  first glance, this result gives a direct confirmation and a
simple physical
interpretation of  both the Hartle-Hawking  wave function of the
Universe {\it
and} the tunneling wave function. However, in all realistic
cosmological
theories, in which $V(\phi)=0$ at its minimum, the Hartle-Hawking
distribution
$\exp\left({3\over 8 V(\phi)}\right)$ is not normalizable. The source
of this
difficulty can be easily
understood: any stationary distribution may exist only due
to  compensation of the classical flow of the field $\phi$
downwards to the minimum of $V(\phi)$ by the diffusion motion
upwards. However, diffusion of the field $\phi$ discussed
above exists only during inflation. Thus, there is no diffusion
motion upwards
from the region $\phi < \phi_e$. Therefore all solutions of  equation
(\ref{E3711}) with the proper  boundary conditions at $\phi = \phi_e$
(i.e. at
the end of inflation)  are non-stationary (decaying) \cite{b19}.

The situation with the probability distribution $P_p$ is much more
interesting
and complicated.  As  was shown in  \cite{b19} its behavior  depends
strongly on
  initial conditions. If  the distribution $P_p$   was initially
concentrated
at $\phi < \phi^*$, where $\phi^*$ is some critical value of the
field,  then
it moves towards small $\phi$ in the same way as $P_c$, i.e. it
cannot become
stationary. On the other hand, if the initial value of the field
$\phi$ is
larger than $\phi^*$, the distribution  moves towards larger and
larger values
of the field $\phi^*$, until it reaches the field $\phi_p$, at which
the
effective potential of the field becomes of the order of Planck
density $M_p^4=1$, where the standard methods of
quantum field
theory in a curved classical space are no longer valid.

Some further steps towards the solution of this problem were made  by
Nambu
and Sasaki  \cite{Nambu} and Miji\'c  \cite{Mijic}. Their
papers
contain many important results and insights. However, Miji\'c  \cite{Mijic}
 did not have a purpose to obtain a complete expression for the
stationary
distribution $P_p(\phi,t|\chi)$\@. The corresponding expressions were
obtained
for various types of potentials $V(\phi)$ in  \cite{Nambu}.
Unfortunately,
according to  \cite{Nambu}, the stationary distribution
$P_p(\phi,t|\chi)$
is almost entirely concentrated at $\phi \gg \phi_p$, i.e. at
$V(\phi) \gg 1$,
where the methods used in  \cite{Nambu} are inapplicable.

We will continue this investigation by writing the system of
stochastic
equations for $P_p$. These equations can be obtained from eqs.
(\ref{b60b}),
(\ref{E3711}) by adding the term $3HP_p$, which appears due to the
growth of
physical volume of the Universe by the factor $1 + 3H(\phi)\, dt$
during each
time interval $dt$ \cite{Nambu,MezhMolch,LLM}:
\begin{equation} \label{b60bx}
\frac{\partial  P_p}{\partial t} =
 \frac{1}{2} \frac{H^{3/2}(\chi)}{2\pi}\, \frac{\partial
}{\partial\chi}
 \left( \frac{H^{3/2}(\chi)}{2\pi} \, \frac{\partial
P_p}{\partial\chi}
\right)
 -  \frac{V'(\chi)}{3H(\chi)} \frac{\partial  P_p}{\partial\chi}
  +3H(\chi)  P_p \ ,
\end{equation}
\begin{equation}\label{E372}
\frac{\partial P_p}{\partial t} =
 \frac{1}{2}  \frac{\partial }{\partial\phi}
 \left( \frac{H^{3/2}(\phi)}{2\pi}\, \frac{\partial }{\partial\phi}
\Bigl(
 \frac{H^{3/2}(\phi)}{2\pi}  P_p \Bigr) +
\frac{V'(\phi)}{3H(\phi)} \, P_p\right)  +  3H(\phi)  P_p\ .
\end{equation}
To find solutions of these equations one must specify boundary
conditions. Behavior of solutions typically is not very sensitive to the
boundary conditions at $\phi_e$; it is sufficient to assume that the diffusion
coefficient (and, correspondingly, the first terms in the r.h.s. of equations
(\ref{b60bx}), (\ref{E372})) vanish for $\phi < \phi_e$.

The
conditions
at the Planck boundary $\phi = \phi_p$  play a more important role.
Investigation of these conditions poses many difficult problems.
First of all,
our diffusion equations are based on the semiclassical approach to
quantum
gravity, which breaks down at  super-Planckian densities. Secondly,
the shape
of the effective potential may be strongly modified by quantum
effects at $\phi
\sim \phi_p$. Finally, our standard interpretation of  the
probability
distribution $P_c$ and $P_p$ breaks down at $V(\phi) > 1$, since at
super-Planckian densities the notion of a classical scalar field in
classical
space-time does not make much sense.

However, these problems by themselves suggest a  possible answer.
Inflation
happens only in  theories with very flat effective potentials. At the
Planck
density nothing can protect the effective potential from becoming
steep.
Hence, one may expect that inflation ceases to exist at $\phi >
\phi_p$, which
leads to the boundary condition
\begin{equation}\label{PlanckBound}
 P_p(\phi_p,t|\chi)= P_p(\phi,t|\chi_p) = 0 \ ,
\end{equation}
where $V(\phi_p) \equiv V(\chi_p) = O(1)$.

There exist some other reasons discussed in \cite{LLM,Vil,BL}, why inflation
and/or self-reproduction of inflationary domains may cease to exist at
super-Planckian densities. We will just assume that this is the case and impose
the boundary conditions (\ref{PlanckBound}) on $P_p$.

One may try to obtain solutions of equations (\ref{b60bx}),
(\ref{E372}) in the
form of the following series of biorthonormal system of
eigenfunctions of the
pair of adjoint linear operators (defined by the left hand sides of
the
equations below):
\begin{equation} \label{eq14}
P_p(\phi,t|\chi) =
\sum_{s=1}^{\infty} { e^{\lambda_s t}\, \psi_s(\chi)\, \pi_s(\phi) }
\ .
\end{equation}
Indeed, this gives us a solution of eq. (\ref{E372}) if
\begin{equation} \label{eq15}
 \frac{1}{2} \frac{H^{3/2}(\chi)}{2\pi} \frac{\partial
}{\partial\chi}
 \left( \frac{H^{3/2}(\chi)}{2\pi} \frac{\partial }{\partial\chi}
\psi_s(\chi) \right)
 - \frac{V'(\chi)}{3H(\chi)} \frac{\partial }{\partial\chi}
\psi_s(\chi)
 + 3H(\chi)  \cdot \psi_s(\chi) =
\lambda_s \, \psi_s(\chi)  \ .
\end{equation}
and\begin{equation} \label{eq17}
\frac{1}{2}  \frac{\partial }{\partial\phi}
 \left( \frac{H^{3/2}(\phi)}{2\pi} \frac{\partial }{\partial\phi}
\left(
 \frac{H^{3/2}(\phi)}{2\pi} \pi_j(\phi) \right) \right)
 + \frac{\partial }{\partial\phi} \left( \frac{V'(\phi)}{3H(\phi)} \,
\pi_j(\phi) \right)
 + 3H(\phi)  \cdot \pi_j(\phi) =
\lambda_j \, \pi_j(\phi)  \ .
\end{equation}
The orthonormality condition reads
\begin{equation} \label{eq20}
\int_{\phi_e}^{\phi_p} { \psi_s(\chi) \, \pi_j(\chi) \, d\chi }
= \delta_{sj} \ .
\end{equation}

In our case (with regular boundary conditions) one can easily show
that the
spectrum of $\lambda_j$ is discrete and bounded from above. Therefore
the
asymptotic solution for $P_p(\phi,t|\chi)$ (in the limit $t
\rightarrow \infty$) is given by
\begin{equation} \label{eq22}
P_p(\phi,t|\chi) = e^{\lambda_1 t}\, \psi_1(\chi) \,
\pi_1(\phi)\, \cdot \left(1 + O\left( e^{-\left(\lambda_1 - \lambda_2
\right) t} \right) \right) \ .
\end{equation}
Here $\psi_1(\chi)$ is the only positive eigenfunction of eq.
(\ref{eq15}),
$\lambda_1$ is the corresponding (real) eigenvalue, and $\pi_1(\phi)$
 is the eigenfunction
of the conjugate operator (\ref{eq17}) with the
same eigenvalue $\lambda_1$\@. Note, that $\lambda_1$ is
the highest eigenvalue, $\mbox{Re} \left( \lambda_1 - \lambda_2
\right) > 0 $\@. This is the reason why the asymptotic equation
(\ref{eq22}) is valid at large $t$\@. We have found \cite{LLM}
that
in realistic theories of inflation a typical time of
relaxing to the asymptotic
regime, $\Delta t \sim (\lambda_1 - \lambda_2)^{-1}$, is extremely
small. Typically it
is only about a few thousands Planck times, i.e. about $10^{-40}$
sec.
 This means that the normalized distribution
\begin{equation} \label{eq22aa}
\tilde{P}_p(\phi,t|\chi) = e^{-\lambda_1 t}
\,P_p(\phi,t|\chi)
\end{equation}
rapidly converges to the time-independent normalized distribution
\begin{equation} \label{eq22a}
\tilde{P}_p(\phi|\chi) \equiv
 \tilde{P}_p(\phi,t \rightarrow \infty|\chi) =  \psi_1(\chi) \,
\pi_1(\phi) \
{}.
\end{equation}
It is this stationary distribution that we were looking for.  Because
the
growing factor  $e^{-\lambda_1 t}$ is the same for all $\phi$ (and
$\chi$), one
can use $\tilde{P}_p$ instead of ${P}_p$ for calculation of all {\it
relative}
probabilities. In particular, $\tilde{P}_p(\phi|\chi)$ gives us the
fraction of
the volume of the Universe occupied by the  field $\phi$, under the
condition
that the corresponding part of the Universe at some time in the past
contained
the field $\chi$.  The
remaining problem is to find the functions $\psi_1(\chi)$ and
$\pi_1(\phi)$,
and to check that all assumptions about the boundary conditions which
we made on the way to eq. (\ref{eq22}) are actually satisfied.

We have solved this problem for chaotic inflation in a wide class of
theories
including the theories with polynomial and exponential effective
potentials
$V(\phi)$ and found the corresponding stationary distributions \cite{LLM}. Here
we will present some of our results for the
theories
${\lambda\over 4}\phi^4$ and $V_o\,e^{\alpha\phi}$.

 Solution of equations (\ref{eq15}) and (\ref{eq17}) for
$\psi_1(\chi)$ and
$\pi_1(\phi)$ in the theory ${\lambda\over 4}\phi^4$  shows that
these
functions are extremely small at  $\phi\sim \phi_e$ and $\chi\sim
\chi_e$. They
grow at large $\phi$ and $\chi$,  then rapidly decrease, and vanish
at $\phi =
\chi=  \phi_p$. With a decrease of $\lambda$ the solutions become
more and more
sharply peaked near the Planck boundary. (The functions $\psi_1$ and
$\pi_1$
for the exponential potential have a similar behavior, but they are
less
sharply peaked near $\phi_p$.) A detailed discussion  of these
solutions will
be contained in \cite{LLM}. The eigenvalues $\lambda_1$
corresponding to
different coupling constants $\lambda$ are given by the following
table:
\begin{center}
\begin{tabular}{|c|c|c|c|c|c|c|c|}
\hline \hline
 $\lambda$ & $1$ & $10^{-1}$ & $10^{-2}$ & $10^{-3}$ & $10^{-4}$ &
$10^{-5}$ &
$10^{-6}$ \\
\hline
 $\lambda_1$ & 2.813 & 4.418 & 5.543 & 6.405 & 7.057 & 7.538 & 7.885
\\
\hline \hline
\end{tabular}
\end{center}
One can find also the second eigenvalue $\lambda_2$. For example, for
$\lambda = 10^{-4}$ one gets  $\lambda_2=6.789$. This means that for
$\lambda =
10^{-4}$ the time of relaxation  to the stationary distribution is
$\Delta t
\sim (\lambda_1-\lambda_2)^{-1} \sim 4 M_p^{-1} \sim 10^{-42}$
seconds --- a
very short time indeed.

Note, that the distribution $\tilde{P}_p(\phi,t|\chi)$ depends on $\phi$ and
$\chi$ very sharply. For example, one can show   that in the theory
${\lambda\over 4}\phi^4$
\begin{equation}\label{SMALLPHI}
\tilde{P}_p(\phi,t|\chi) \sim  \psi_1(\chi)\,   \phi^{\sqrt{6\pi\over
\lambda}\lambda_1} \,   \ .
\end{equation}
for a sufficiently small $\phi$ (for $\phi < \lambda^{-1/8}$) This is an
extremely sharp dependence indeed. For example, for the realistic value
$\lambda \sim 10^{-13}$
one obtains $\tilde{P}_p(\phi,\tau|\chi) \sim  \psi_1(\chi) \, \phi^{10^8}$   !
This  result will be important for us later, when we will be discussing
predictions of quantum cosmology for the value of the cosmological parameter
$\Omega$ \cite{LLM2}.

Note that the parameter $\lambda_1$ shows the speed of exponential
expansion of
the volume filled by a given field $\phi$. {\it This speed does not
depend on
the field $\phi$}, and has the same order of magnitude as the speed
of
expansion at the Planck density. Indeed, $\lambda_1$ should be
compared to
$3H(\phi)= 2\sqrt{6\pi V(\phi)}$, which is  equal to $2\sqrt{6\pi}\approx 8.68$
at the
Planck density. It can be shown\cite{LLM} that  in the limit
$\lambda
\to 0$ the eigenvalue $\lambda_1$ also becomes equal to
$2\sqrt{6\pi} \approx
8.681$. The meaning of this result is very simple: in the limit
$\lambda \to 0$
our solution becomes completely concentrated near the Planck
boundary, and
$\lambda_1$ becomes equal to $3H(\phi_p)$.

One should emphasize    that the factor $e^{\lambda_1 t}$
  gives the rate of growth of the combined volume
of all domains with a given field $\phi$ (or of all domains
containing matter with a given density) {\it  not only at very
large $\phi$, where quantum fluctuations are large, but at small
$\phi$ as well, and even after inflation} \cite{LLM}. This
result may seem absolutely unexpected, since the volume of each
particular inflationary domain grows like $e^{3H(\phi)t}$, and
after inflation the law of expansion becomes completely
different.  One should distinguish, however, between the growth
of each particular domain, accompanied by a decrease of density
inside it, and the growth of the total volume of all domains
containing matter with a given (constant) density. In the
standard big bang theory the second possibility did not exist,
since the energy density was assumed to be  the same in all
parts of the Universe (``cosmological principle''), and it was
not constant in time.

The reason why there is a universal expansion rate $e^{\lambda_1 t}$
can be understood as follows. Because of the self-reproduction
of the Universe there always exist many domains with $\phi \sim
\phi_{p}$,  and their combined volume grows almost as fast as
$e^{3H(\phi_p) t}$. Then the field $\phi$ inside some of these
domains decreases. The  total volume of domains containing some
small field $\phi$ grows not only due to expansion $\sim
e^{3H(\phi_p) t}$, but mainly due to the unceasing process of
expansion of domains with large $\phi$ and their subsequent
rolling (or diffusion) towards small $\phi$.

But what  about the field $\phi$ which is already at the Planck
boundary? Why
do the corresponding domains not  grow exactly with the Planckian
Hubble
constant $H(\phi_p) = 2\sqrt{6\pi}/3$\,? It happens partially due to
diffusion
and slow rolling of the field towards smaller $\phi$. However, the
leading
effect  is the destructive diffusion towards the space-time foam with
$\phi >
\phi_p$. One may  visualize this process  by painting white all
domains with
$V(\phi) < 1$, and by painting black domains filled by space-time
foam with
$V(\phi) > 1$.  Then each time $H^{-1}(\phi_p)$ the volume of white
domains
with $\phi \sim \phi_p$ grows approximately $e^3$ times, but some
`black holes'
appear in these domains, and, as a result, the total volume of white
domains
increases only $e^{3\lambda_1/2\sqrt{6\pi}}$ times. This suggests (by
analogy
with \cite{ArVil})   calling the factor
$d_f = 3\lambda_1/2\sqrt{6\pi}$
`the fractal dimension of classical space-time', or  `the fractal
dimension of
the inflationary Universe'. However, one should keep in mind that the fractal
structure of the
inflationary
Universe in the chaotic inflation scenario in general is more
complicated than
in the new or old inflation and cannot be completely specified just
by one
fractal dimension \cite{LLM}.

The distribution $\tilde{P}_p(\phi | \chi)  =  \psi_1(\chi) \,
\pi_1(\phi)$
which we have obtained  does not depend on time $t$. However, in
general
relativity one may use many different time parametrizations, and the
same
physics can be described
differently in different `times'. One of the most natural choices of
time in
the context of stochastic approach to inflation is the time $\tau  =
\ln{{a\left(x, t \right) \over a(x,0)}} =  \int{H(\phi(x,t),t)\ dt}$
\cite{b60,Bond}.  Here $a\left(x, t \right)$ is a local value of the
scale
factor in the inflationary Universe. By using this time variable, we
were able
to obtain not only numerical solutions to the stochastic equations,
but also
simple asymptotic expressions describing these solutions. For
example, for the
theory ${\lambda\over 4 } \phi^4$ both the eigenvalue $\lambda_1$ and
the `fractal dimension' $d_f$ (which in this case refers both to the
Planck
boundary at $\phi_p$ and to the end of inflation at $\phi_e$) are
given by $d_f
= \lambda_1 \sim 3-1.1\, \sqrt \lambda$, and the stationary
distribution is
\begin{eqnarray}
\tilde{P}_p(\phi,\tau|\chi) &\sim &  \exp\Bigl(-{3\over
8V(\chi)}\Bigr)\,
\Bigl({1\over V(\chi)+0.4} - {1\over1.4}\Bigr)\, \cdot \,   \phi
\,\exp\Bigl(-{\pi\, (3-\lambda_1)\phi^2}\Bigr) \nonumber \\
 &\sim & \exp\Bigl(-{3\over 2 \lambda\, \chi^4}\Bigr)\, \Bigl({4\over
\lambda
\chi^4+1.6} - {1\over1.4}\Bigr)\, \cdot \,   \phi \, \exp\Bigl(- 3.5
\sqrt\lambda\phi^2\Bigr)\ .
\end{eqnarray}
Note that the first factor coincides with the square of the tunneling
wave
function  \cite{Creation}! This expression is valid  in the whole
interval from
$\phi_e$ to  $\phi_p$ and it correctly describes asymptotic behavior
of
$\tilde{P}_p(\phi,\tau|\chi)$ both at  $\chi \sim \chi_e$ and at
$\chi \sim
\chi_p$.

A similar investigation can be carried out for the theory $V(\phi) =
V_o\
e^{\alpha\phi}$. The corresponding solution is
\begin{eqnarray}
\tilde{P}_p(\phi,\tau|\chi) &\sim &  \exp\Bigl(-{3\over
8V(\chi)}\Bigr)\,
\Bigl({1\over V(\chi)} - {1}\Bigr)\, \cdot \,  \Bigl({1\over V(\phi)}
-
{1}\Bigr)\, V^{-1/2}(\phi)\ .
\end{eqnarray}
This expression gives a  rather good approximation for
$\tilde{P}_p(\phi,\tau|\chi)$ for all $\phi$ and  $\chi$.

The main result of our work is that under certain conditions the
properties of
our   Universe  can be described by a time-independent probability
distribution, which we have found for theories with polynomial and
exponential
effective potentials. A lot of work still has to be done to verify
this
conclusion, see \cite{LLM}. However, once this result is taken
seriously, one should consider its interpretation and  rather unusual
implications.

When  making cosmological observations, we study our part of the
Universe and
find that in this part inflation ended  about $t_e \sim 10^{10}$
years ago. The
standard assumption of the first models of inflation was that the
total
duration of the inflationary stage was  $\Delta t \sim 10^{-35}$
seconds. Thus
one could come to an obvious conclusion that our part of the Universe
was
created in the big bang, at the time  $t_e + \Delta t \sim 10^{10}$
years ago.
However, in our scenario the answer is quite different.

Let us consider an inflationary domain which gave rise to the process
of
self-reproduction of new inflationary domains. For illustrative
purposes, one
can visualize self-reproduction of inflationary domains  as a
branching
process, which gives a qualitatively correct description of  the
actual
physical process we consider. During this process, the first
inflationary
domain of  initial radius $\sim H^{-1}(\phi)$ within the time
$H^{-1}(\phi)$
splits into $e^{3} \sim 20$ independent inflationary domains of
similar size.
Each of them contains a slightly different field $\phi$,  modified
both by
classical motion down to the minimum of $V(\phi)$ and by
long-wavelength
quantum fluctuations of amplitude $\sim H/2\pi$. After the next time
step
$H^{-1}(\phi)$, which will be slightly different for each of these
domains,
they split again, etc. The whole process now looks like a branching
tree
growing from the first (root) domain. The radius of each branch is
given by
$H^{-1}$;  the total volume of all domains at any given time $t$
corresponds to
the `cross-section' of all branches of the tree at that time, and is
proportional to the number of branches. This volume  rapidly grows,
but when
calculating it, one should   take into account  that those branches,
in which
the field becomes  larger than $\phi_p$,  die and fall down from the
tree, and
each branch in which the field becomes  smaller than  $\phi_e$, ends
on an
apple (a part of the Universe where inflation ended and life became
possible).

One of our results is that even after we discard at each given moment
the dead
branches and the branches ended with apples, the total volume of live
(inflationary) domains will continue growing exponentially, as
$e^{\lambda_1
t}$. What is even more interesting, we have found that very soon the
portion of
branches with given properties (with given values of scalar fields,
etc.)
becomes time-independent. Thus, by observing any finite part of a
tree  at any
given time $t$ one cannot tell how old   the tree is.

To give a most dramatic representation of our conclusions, let us see
where
most of the apples  grow. This can be done simply by integrating
$e^{\lambda_1
t}$ from $t=0$ to $t = T$ and taking the limit as $T\to \infty$. The
result
obviously diverges at large $T$ as
 $\lambda_1^{-1}\, e^{\lambda_1 T}$, which
means that most  apples grow at an indefinitely large distance from
the root.
In other words, if we ask what is the total duration of inflation
which
produced a typical apple, the answer is that it is indefinitely
large.

This conclusion may seem very strange. Indeed, if one takes a typical
point in
the root domain, one can show that inflation at this point ends
within a finite
time $\Delta t \sim 10^{-35}$ seconds. This is a correct (though
model-dependent) result which can be confirmed by stochastic methods,
using the
distribution $P_c(\phi,\Delta t|\chi)$ \cite{b19}.  How could  it
happen that
the duration of inflation was any longer than $10^{-35}$ seconds?

The answer is related to the choice between $P_c$ and $P_p$, or
between roots
and fruits. Typical points in the root domain drop out from the
process of
inflation within $10^{-35}$ seconds. The number of those points which
drop out
from inflation at a much later stage is exponentially suppressed,
but they
produce the main part of the total volume of the Universe.
Note that the length of each particular branch continued back in time
may well
be finite \cite{Vil}. However, there is no upper limit to the length
of each
branch, and, as we have seen, the longest branches produce almost all
parts of
the Universe with properties similar to the properties of the part
where we
live now. Since by local observations we can tell nothing about our
distance in
time from the root domain, our probabilistic arguments suggest that
the root
domain is, perhaps, indefinitely far away from us. Moreover, nothing
in our
part of the Universe depends on the distance from the root domain,
and,
consequently, on the distance from the big bang.

Thus,  inflation  solves many problems of the big bang theory and
ensures that
this theory provides an excellent description of the local structure
of the
Universe. However,  after making all kinds of improvements of  this
theory, we
are now winding up with a model of a stationary Universe, in
which the
notion of the big bang    loses its dominant position, being removed
to the indefinite past.

\

\section{Cosmological constant, $\Omega <1$, and all that }
When inflationary theory was first formulated, we did not know
how much  it was going to influence our  understanding of the structure of the
Universe.  We were happy   that inflation provided an easy explanation of the
homogeneity of the Universe. However, we did not know that the same theory
simultaneously predicts that on the extremely large scale the Universe becomes
entirely inhomogeneous, and that this inhomogeneity is good, since it is one of
the manifestations of the process of self-reproduction of inflationary
universe.

The new picture of the Universe which emerges now is very unusual, and we are
still in the process of learning how to ask proper questions in the context of
the new cosmological paradigm. Previously we assumed  that we live in a
Universe which has the same properties everywhere (``cosmological principle'').
 Then one could make a guess about the most natural initial conditions in the
Universe, and all the rest followed almost automatically. Now we learned that
even if one begins with a non-uniform Universe, later it   becomes extremely
homogeneous on a very large scale. However, simultaneously it becomes
absolutely non-uniform on a much greater scale. The Universe becomes divided
into different exponentially large regions where the laws of low-energy physics
become different. In certain cases the relative fraction of volume of the
Universe in a state with given fields or with a given density does not depend
on time, whereas the total volume of all parts of the Universe continues
growing exponentially.

In this situation the concept of naturalness becomes much more ambiguous than
before, the problem of introduction of a proper measure of probability becomes
most important.  One of the most natural choices of such measure is given by
the probability distribution $P_p(\phi,t|\chi)$. The hypothesis behind this
proposal is that we are typical, and therefore we live in those parts of the
Universe where   most other people do.  The total number of people which can
live in domains with given properties should be proportional to the total
volume of these domains. If $P_p(\phi,t|\chi)$ is stationary, then it seems
reasonable to use it as a measure of the total volume of these domains at any
given moment of time $t$. A detailed discussion of this distribution and all
possible problems involved is contained in \cite{LLM}. Some possible
applications of this probability distribution for finding the most probable
values of the effective gravitational constant in the Brans-Dicke inflationary
cosmology can be found in \cite{GBLL}. We have found there that inflation in
the Brans-Dicke theory leads to division of the Universe into different
exponentially large domains with different values of the gravitational
constant, and, correspondingly, with different values of density perturbations.
We have shown also that the probability distribution $P_p(\phi,t|\chi)$ in
certain versions of this theory can be non-stationary, running away towards
infinitely large $\phi$, corresponding to infinitely large values of the
effective Planck mass. To obtain a correct interpretation of these results one
should develop some combination of stochastic approach to inflation and
anthropic principle.

A step in this direction was made in our recent paper with Juan
Garc\'{\i}a--Bellido \cite{BL}, where  we investigated possible consequences of
stochastic effects during inflation in application to the cosmological constant
problem.  We have shown, in particular, that by combining the baby universe
theory with the Starobinsky model one can solve the problem of cosmological
constant. The idea is that the
overall growth  of the total volume of all domains of the  Universe containing
matter with the density $\rho \sim 10^{-29}$ g$\cdot$cm$^{-3}$ is proportional
to $e^{\lambda_1 t}$, where $\lambda_1$ is a parameter which depends on other
constants of the theory, including the cosmological constant. As we have shown
in \cite{BL}, in the context of the Starobinsky model this rate increases as
the cosmological constant becomes smaller. However, this solution works only if
one believes into the baby universe theory (which allows one to compare
universes with different values of the cosmological constant), and only if the
cosmological constant cannot become negative. Thus, it cannot be considered as
 a final solution of the cosmological constant problem, but at least we have a
completely new way to address this issue.

In what follows
I will briefly describe some    nonperturbative effects  which may lead to a
considerable local deviation of density from the critical density of a flat
universe.   Results which I am going to discuss are very recent; they have been
obtained in a collaboration  with Dmitri Linde and Arthur Mezhlumian
\cite{LLM2}.

Let us consider all parts of inflationary universe which contain a given field
$\phi$ at a given moment of time $t$. One may wonder, what was the value of
this field in those domains at the moment $t - H^{-1}$ ? The  answer is simple:
One should add to $\phi$ the value of its classical drift $\Delta \phi$ during
the time $H^{-1}$,  $\Delta \phi = \dot\phi H^{-1}$. One should also add the
amplitude of a quantum jump $\delta \phi$. The typical   jump  is given by
$\delta \phi = \pm {H\over 2\pi}$. At the last stages  of inflation this
quantity is by many orders of magnitude smaller than $\Delta \phi$. However, in
which sense   jumps $\pm {H\over 2\pi}$ are typical? If we consider any
particular initial value of the field $\phi$, then the typical jump from this
point is indeed given by $\pm {H\over 2\pi}$. However, if we are considering
all domains with a given $\phi$ and trying to find all those domains from which
the field $\phi$ could originate back in time, the answer may be quite
different. Indeed,   the total volume of all domains with a given field $\phi$
at any moment of time $t$ strongly depends on $\phi$:   \,$\tilde{P}_p(\phi)
\sim   \phi^{\sqrt{6\pi\over \lambda}\lambda_1} \sim \phi^{10^8}$, see eq.
(\ref{SMALLPHI}). This means that the total volume of all domains which could
jump towards the given field $\phi$ from the value $\phi +\delta \phi$ will be
enhanced by a large  additional factor $ {\tilde{P}_p(\phi +\delta \phi)\over
\tilde{P}_p(\phi)} \sim   \Bigl(1+{\delta\phi\over \phi}\Bigr)^{\sqrt{6\pi\over
\lambda}\lambda_1}$. On the other hand, the probability of large jumps
$\delta\phi$ is suppressed by the Gaussian factor
$\exp\Bigl(-{2\pi^2\delta\phi^2\over H^2}\Bigr)$.
One can easily verify that the product of these two factors has a sharp maximum
at $\delta\phi = \lambda_1 \phi   \cdot {H\over 2\pi}$, and the width of this
maximum is of the order ${H\over 2\pi}$. In other words, most of the domains of
a given field $\phi$ are formed due to   jumps which are greater than the
``typical'' ones by a factor $\lambda_1 \phi \pm O(1) $.

Our part of the Universe in the inflationary scenario with $V(\phi) =
{\lambda\over 4} \phi^4$ is formed at $\phi~\sim~5$ (in the units $M_p = 1$),
and the constant $\lambda_1 \approx   2\sqrt{6\pi} \sim 8.68$ for our choice of
boundary conditions. This means that our part of the  Universe should be
created as a result of a jump down which is about  $\lambda_1 \phi \sim 40$
times greater than the standard jump. The standard jumps   lead to   density
perturbations of the   amplitude $\delta\rho \sim 5\cdot10^{-5} \rho_c$ (in the
normalization of \cite{MyBook}). Thus, according to our nonperturbative
analysis, we should live inside a region where density is smaller than the
critical density by about $\delta\rho \sim 2\cdot10^{-3} \rho_c$. As we already
mentioned, the probability of such fluctuations should be suppressed  by
$\exp\Bigl(-{2\pi^2\delta\phi^2\over H^2}\Bigr)$, which in our case gives the
suppression factor $ \sim  \exp(-10^3)$. It is well known that exponentially
suppressed perturbations typically give rise to spherically symmetric bubbles.
Note also, that the Gaussian distribution suppressing the amplitude of the
perturbations refers to the amplitude of a perturbation in its maximum. It is
possible that we live not in the place corresponding to the maximum of the
fluctuation. However, this could only happen   if the nonperturbative jump down
was even greater in the amplitude that we expected. Meanwhile, as we already
mentioned, the distribution of the amplitudes of such jumps has  width of only
about ${H\over 2\pi}$. This means that we should live very close to the center
of the giant fluctuation, and the difference of energy densities between the
place where we live and the center of the ``bubble'' should be only about the
same amplitude as the typical perturbative fluctuation  $\delta\rho \sim
5\cdot10^{-5} \rho_c$. In other words, we should live very close to the  center
of the nearly perfect spherically symmetric bubble, which contains matter with
a smaller energy density than the matter outside it.

It is very tempting to interpret this effect in such a way that the Universe
around us becomes locally open, with $1 - \Omega  \sim  10^{-3}$.  The true
description of this effect is, of course, much more complicated, perhaps we
should see  the Hubble constant slightly growing at large distances. Note that
in the simplest theory which we considered the magnitude of this effect is
rather small. However, this smallness is proportional to the coefficient
$\lambda_1$, which is related to the maximal value of the Hubble constant
during inflation. In some inflationary models based  on the Brans-Dicke
theories  the effective value of $\lambda_1$ may be extremely large or even
infinite \cite{GBLL,BL}. Thus, the typical deviation of density of the Universe
from the average value in certain models can be very significant. If this is
correct, then the degree of the local decrease of density can be used to
determine the value of the parameter $\lambda_1$. This means that by studying
the large-scale structure of the Universe one can get   information about the
properties of space-time at Planckian densities!

This effect is extremely unusual. We became partially satisfied by our
understanding of this effect only after we confirmed its existence   by four
different methods, including computer simulations \cite{LLM2}. However, there
are still many problems associated with it. We are still trying to verify the
existence of this effect  and to find an adequate interpretation of our
results.  The most important problem is related to the choice of measure in
inflationary cosmology.  If we  repeat our investigation using instead of
$P_p(\phi,t)$ the distribution $P_p(\phi, \tau)$, where $\tau \sim \log a$, we
do not get any nonperturbative effects of that kind. The source of this
ambiguity can be easily understood. The total volume of all parts of an
eternally inflating universe is infinite in the limit $t \to \infty$ (or $\tau
\to \infty$). Therefore when we are trying to compare volumes of domains with
different properties, we are comparing infinities. This leads to answers
depending on the way we are making this comparison, see the discussion of this
issue in \cite{LLM}--\cite{BL}.

Until we  resolve the problem of measure in quantum cosmology, it will be hard
to prove that the effect discussed above should actually take place. On the
other hand, it is fair to say that now we cannot exclude the possibility of
such an effect, and this by itself  is a very unexpected conclusion. Few years
ago we would say that such effects definitely contradict   inflationary
cosmology. Now we can only say that  this is an open  question to be studied
experimentally, and that maybe cosmological observations can tell us something
new about the nature of quantum gravity.

\section{Conclusions}

 The first realistic versions of inflationary cosmology have been proposed more
than ten years ago. Since that time inflationary theory gradually became a very
popular tool for investigation of the physical processes in the   early
universe. It also provides us with a good framework to study the large-scale
structure of the Universe. However, inflationary theory is still at the stage
of its active development. Within the last one or two years we have realized
that inflation may end not only by slow rolling or by bubble formation, but in
a kind of waterfall regime, associated with hybrid inflation. We have learned
that reheating after inflation occurs in a much more complicated way than  we
thought before.   I expect the theory of reheating to become a subject of
intensive study in the near future. We have also learned that our understanding
of the way inflation solves the monopole problem was absolutely incomplete,
since we neglected the possibility that topological defects can exponentially
grow. Finally, it was realized that inflationary theory may give rise to the
theory of a stationary universe. This development completely changes our
understanding of the structure  and the fate of our Universe, and of our own
place in the world.
  It is amazing that after  many years of its active development inflationary
theory still reveals so many surprising features and    provides us with so
many challenging problems to be solved!

\

\subsection*{Acknowledgements}
These lectures were based on the recent results which have been obtained in a
collaboration with  J.  Garc\'{\i}a--Bellido,   L.A. Kofman, D.A.  Linde, A.
Mezhlumian  and A.A. Starobinsky.  I should express my gratitude not only to my
co-authors, but also  to  R.
Brandenberger, D. Boyanovsky, H.J. de Vega, R. Holman, D.-S. Lee, V. Mukhanov,
Y. Shtanov,    A. Singh, J. Traschen, and A. Vilenkin
 for many valuable discussions. I appreciate warm hospitality of the Organizers
of   the School  on Particle Physics and Cosmology at Lake Louise, Canada,  of
the Marcel Grossmann Conference, Stanford,  of the Workshop on Birth of the
Universe, Roma,   of the Symposium on Elementary Particle Physics, Capri, and
of the School of Astrophysics,  Erice, Italy.  This work  was supported in part
by NSF
grant PHY-8612280.

\end{document}